\documentstyle[preprint,aps,prd,epsfig]{revtex}

\def\lprox{\mathrel{\raise .3ex\hbox{$<$\kern-
.75em\lower1ex\hbox{$\sim$}}}}

\def\gprox{\mathrel{\raise .3ex\hbox{$>$\kern-
.75em\lower1ex\hbox{$\sim$}}}}

\def\be{\begin{equation}}
\def\ee{\end{equation}}	
\def\ben{\begin{eqnarray}}
\def\een{\end{eqnarray}}

\begin{document}

\draft

\title{\bf Quantum lightcone fluctuations in theories with  extra dimensions}

\author{Hongwei Yu \footnote{e-mail: hwyu@cosmos2.phy.tufts.edu} 
        and L. H. Ford \footnote{e-mail: ford@cosmos2.phy.tufts.edu}}

\address{Institute of Cosmology, Department of Physics and Astronomy\\
	Tufts University, Medford, MA 02155, USA}

\date{\today}

\maketitle

\tightenlines

\begin{abstract}
The effects of small extra dimensions upon quantum fluctuations of the
lightcone are examined. We argue that compactified extra dimensions 
modify the quantum fluctuations of gravitational field so as to induce
lightcone fluctuations. This phenomenon can be viewed as being related
to the Casimir effect. The observable manifestation of the lightcone 
fluctuations is broadening of spectral lines from distant sources. In 
this paper, we further develop the formalism used to describe the lightcone 
fluctuations, and then perform explicit calculations for several models
with flat extra dimensions. In the case of one extra compactified dimension,
we find a large effect which places severe constraints on such models.
When there is more than one compactified dimension, the effect is much weaker
and does not place a meaningful constraint. We also discuss some brane
worlds scenarios, in which gravitons satisfy Dirichlet or Neumann
boundary conditions on parallel four-dimensional branes, separated by
one or more flat extra dimensions.
\end{abstract}



\section{ Introduction}
 
One of the most challenging problems in modern physics is the unification of the gravitational 
interaction with other known interactions in nature. Many attempts, from early 
Kaluza-Klein
 theory~\cite{ACF} to present
supergravity and superstring theories, all 
 involve going to higher dimensions and postulating the existence of extra 
spatial dimensions. The modern view regarding 
these postulated extra dimensions is that they are a physical reality, 
rather than merely a technical intermediate device for
obtaining rather complicated four-dimensional theories from simpler Lagrangians 
in higher dimensions. 
If these  extra dimensions really exist, one must explain why they are not seen.
 The usual 
answer is that they curl into an extremely small compactified manifold, 
possibly   as small as the Planck length scale, 
 $l_{pl}=1.6\times 10^{-33}$ cm. Therefore low-energy physics should be 
insensitive to them until distances of the compactification scale are being 
probed. In general, one has the possibility of 
observing the presence of the 
extra dimensions in a scattering experiment in which energies greater than that
associated with the compactification scale are achieved. Various upper bounds
have been put on the size of possible extra dimensions \cite{ANT90,KS91,AB94}. 
 For example, an upper bound of $\sim$ 1 Tev was given in orbifold  
compactifications of superstrings \cite{AB94}. However, if only gravity
propagates in the extra dimensions, the upper bound can be much larger.  
A recent proposal  is that the fundamental scale of quantum 
gravity  can be as low as few Tev and the observed weakness of gravity is the 
result of large extra dimensions in which only gravity can propagate \cite{ADD}.
 This scenario could be realized in the context of
several string models \cite{WLST} in which one has a set of three-branes 
(3+1 dimensional 
spacetime) in the entire spacetime with extra dimensions.  The Standard Model 
particles are confined to one of the branes,  
 while gravitons propagate freely in the
entire bulk.  The  size of extra
dimensions could then be as large as  1 mm in this type of model. 
Extra dimensions of sufficiently large size may manifest themselves in 
particle colliders\cite{Colliders}  and in the possible deviation from Newton's law at short 
distances\cite{DN}, and they may also have  implications in gauge 
unification\cite{GU} and cosmology\cite{Cosmo}.

However, a question arises naturally as to whether there are any lower bounds on the sizes of extra dimensions. 
 It is the common belief that the existence of extra dimensions has no effect on low-energy physics
as long as  they are extremely small. Recently, 
we have argued that this is not 
 the case, because of lightcone fluctuations arising 
from the quantum
gravitational vacuum fluctuations due to compactification of spatial 
dimensions~\cite{YUF,YUF2}.  An explicit calculation was carried out in the five-dimensional
prototypical Kaluza-Klein model which showed that the periodic  compactification of  the extra 
spatial 
dimensions gives rise to stochastic fluctuations in the apparent speed of light 
which grow as the compactification scale decreases and  are in principle observable. 
 Basically,  the smaller the size of the 
compactified dimensions , the larger are the fluctuations that  result. 
This is closely related to the Casimir effect,  the vacuum energy occurring 
whenever boundary conditions are imposed on a quantum
field.  The gravitational Casimir energy in the five-dimensional case with one compactified 
spatial dimension was studied in \cite{APCH}, where a nonzero energy density was found, which
tends to make the extra dimension  contract. This raises the question of stability of the extra
dimensions. It is possible, however, that the Casimir energy arising from the quantum 
gravitational field and other matter fields may be made to cancel each other~\cite{RRT}, thus stabilizing the extra dimensions. Quantum lightcone fluctuations  due to the compactification of
 spatial dimensions~\cite{YUF,YUF2,Ford95}, although similar in nature to the Casimir effect, come solely from 
gravitons. Hence, no similar cancellation is to be expected.  

In an earlier work \cite{YUF2}, we examined lightcone fluctuation in a 
five-dimensional model with periodic compactification. We found that there
seem to be observable effects which essentially rule out this model. That is,
we derived a lower bound on the compactification scale which is larger than
upper bounds derived from other considerations. The purpose of the present paper
is threefold: to further develop the basic formalism, to provide more details of
the five-dimensional calculation, and to extend the analysis to some higher
dimensional models.

In Section II, we give a brief review of the formalism, discuss the observability of 
lightcone fluctuations, and derive the graviton two-point functions in the 
transverse tracefree gauge for 
spacetimes with an arbitrary number of dimensions (detailed calculations will 
given in the Appendix). 
In Section III, we examine light cone fluctuations in spacetimes with extra 
dimensions  periodically compactified into a torus. The five-dimensional 
prototypical Kaluza-Klein 
model, of which some results have already been reported,  will be studied in 
great detail.
Higher dimensions up through 11 will also be discussed.  
Section IV deals with light cone fluctuations in the brane-world scenario, as 
motivated 
by a recent proposal of extra dimensions of macroscopic size.  We will summarize 
and conclude in Section V.   


\section{Observability of light cone fluctuations and the graviton 
two-point function in the TT gauge}


  To begin, let us examine a $d=4+n$ dimensional spacetime with $n$  extra dimensions. 
 Consider a flat background  spacetime  with a linearized perturbation 
$h_{\mu\nu}$ propagating upon it , so the spacetime metric
may be written as  $
ds^2  = (\eta_{\mu\nu} +h_{\mu\nu})dx^\mu dx^\nu
= dt^2 -d{\bf x}^2 + h_{\mu\nu}dx^\mu dx^\nu \, ,  $
where the indices $\mu,\nu$ run through $0,1,2,3,...,3+n$.
Let $\sigma(x,x')$ be one half of the squared geodesic distance between
 a pair of spacetime points $x$ and $x'$,   and $\sigma_0(x,x')$ 
be the corresponding quantity in the flat background . 
In the presence of a linearized metric  perturbation,
 $h_{\mu\nu}$, we may expand $
\sigma = \sigma_0 + \sigma_1 + O(h^2_{\mu\nu}) \, .$
Here $\sigma_1$ is first order  in $h_{\mu\nu}$. 
If we quantize  $h_{\mu\nu}$,
then  quantum gravitational vacuum
fluctuations will  lead to fluctuations in the geodesic separation, and therefore induce
 lightcone fluctuations.  In particular, we have $\langle \sigma_1^2 \rangle \not= 0$, since $\sigma_1$  becomes a quantum operator when the metric perturbations are quantized.   The quantum lightcone fluctuations give rise to 
stochastic fluctuations in the speed of light, which may produce an observable 
time delay or advance $\Delta t$ in the arrival times of pulses.


\subsection{Observability of light cone fluctuation }


Here, we shall discuss how light cone fluctuations characterized by
 $\langle \sigma_1^2 \rangle$ are related to physical observable quantities. For 
this purpose, 
let us  consider the propagation of light pulses
between a source and a detector separated by a distance $r$ on a flat background with 
quantized linear perturbations. For a pulse which is delayed or advanced by time $\Delta t$,
which is much less than $r$, one finds
\be
\sigma=\sigma_0+\sigma_1+....={1\over2}[(r+\Delta t)^2-r^2]\approx r\Delta t\,.
\label{eq:separation}
\ee
Square the above equation and take the average over a given  quantum state of gravitons 
$|\phi\rangle$ (e.g. the vacuum states associated with  compactification of spatial dimensions ), 
\be
\Delta t_{\phi}^2={\langle \phi| \sigma_1^2 |\phi\rangle\over r^2}\,.
\label{eq:TPHI}
\ee
This result is, however,  divergent due to the formal divergence of
 $\langle\phi| \sigma_1^2 |\phi\rangle $. One can 
 define 
 an observable $\Delta t_{obs}$  by subtracting from Eq.~(\ref{eq:TPHI}) the corresponding 
quantity, $\Delta t_0^2$, for the  vacuum state as follows
\be
 \Delta t_{obs}^2=|\Delta t_{\phi}^2-\Delta t_0^2 |=
{|\langle \phi| \sigma_1^2 |\phi\rangle-\langle 0| \sigma_1^2 |0\rangle|\over r^2}\equiv
{|\langle \sigma_1^2 \rangle_R|\over r}\,.
\ee
Here we  take the absolute value of the difference between 
$\Delta t_{\phi}^2$ and $\Delta t_0^2$, because the observable quantity $\Delta t_{obs}^2$
has to be  a positive real number. Note that we can also  get this result from the gravitational 
quantum average of the retarded Green's function $\langle G_{ret}(x,x')\rangle$ when 
$\langle \sigma_1^2 \rangle_R > 0$\cite{Ford95}.  
Therefore, the root-mean-squared  deviation from the classical propagation time is given by
\be
\Delta t_{obs}= {\sqrt{|\langle \sigma_1^2 \rangle_R|}\over r}\,.
\label{eq:MDT}
\ee

At this point, a question may arise as to whether  the formal procedure of 
taking the absolute value in
 deriving the relation between $\Delta t$ and $\langle \sigma_1^2 \rangle_R$, 
Eq.~(\ref{eq:MDT}),  is a
 reasonable one, or whether a meaningful relation between $\Delta t$ and 
$\langle \sigma_1^2 \rangle_R$ can be 
established only when $\langle \sigma_1^2 \rangle_R > 0$. We shall argue that 
a result essentially
the same as  Eq.~(\ref{eq:MDT}) can be obtained by the following analysis which
 avoids the sign problem.
Instead of squaring the Eq.~(\ref{eq:separation}), we  take the fourth 
power of both sides and average over a quantum 
vacuum state to yield
\be
\Delta t_{\phi}^4={\langle \phi| \sigma_1^4 |\phi\rangle\over r^4}\,.
\label{eq:TPHI4}
\ee
 We can regularize $\langle \phi| \sigma_1^4 |\phi\rangle$ by normal ordering
 and define
\be
\langle \phi| \sigma_1^4 |\phi\rangle_R=
\langle \phi| :\sigma_1^4 :|\phi\rangle\,.
\ee
For a free field $\psi$ and a quantum vacuum state one can show by use of 
Wick's theorem that 
\be 
\langle  :\psi^4 :\rangle=3\langle  :\psi^2 :\rangle^2=3\langle \psi^2 \rangle_R ^2\,. 
\ee
Therefore one  finds that 
\be
\Delta t_{obs}= {3^{1/4}\sqrt{|\langle \sigma_1^2 \rangle_R|}\over r}
\approx {\sqrt{|\langle \sigma_1^2 \rangle_R|}\over r}\,.
\ee
This is essentially the same as Eq.~(\ref{eq:MDT}), apart from a dimensionless 
factor of order unity.
We should note that there may be other ways to define the quartic operator,
$\sigma_1^4$. One possibility is to let $\sigma_1^4 = (:\sigma_1^2:)^2$,
provided that the integrals involved can be defined. Both of these definitions
were discussed in Ref.~\cite{WF99}. There it was found  in some model
cases that the two definitions yield the same result, apart from numerical
factors of order unity, which are not important for the present purposes.

Note that $\Delta t$ is the ensemble averaged deviation, not 
necessarily the expected variation in
flight time, $\delta t $,  of two pulses emitted close together in time. 
The latter is given by   $\Delta t$ only when the correlation time between 
successive pulses is less than the time separation of the pulses. This can be 
understood physically as due 
to the fact that the gravitational field may not fluctuate significantly in the
 interval between the two pulses.
This point is discussed in detail in Ref.~\cite{Ford96}.
These stochastic fluctuations in the apparent velocity of light arising from
 quantum gravitational 
fluctuations are in principle observable, since 
they may lead to a spread in the arrival times of pulses from
distant astrophysical sources, or the broadening of the spectral lines. 
Lightcone fluctuations and their possible astrophysical observability have 
been recently discussed in a somewhat different framework 
in Refs.~\cite{AEMN,EMN}.


\subsection{An Alternative Derivation of $\Delta t$}
\label{sec:del_t}

  In order to find $\Delta t$ in a particular situation, 
we need to calculate the quantum expectation value 
 $\langle \sigma_1^2 \rangle_R$ in any 
chosen quantum state $|\psi\rangle$, which can be shown to be given by 
 \footnote{ Although the derivations there were given in 3+1 
dimensions, the generalization to arbitrary dimensions is 
straightforward.}\cite{Ford95,YUF}
\be
\langle \sigma_1^2 \rangle_R ={1\over 8}(\Delta r)^2
\int_{r_0}^{r_1} dr \int_{r_0}^{r_1} dr'
\:\,  n^{\mu} n^{\nu} n^{\rho} n^{\sigma}
\:\, G^{R}_{\mu\nu\rho\sigma}(x,x') \,.
\label{eq:interval}
\ee
Here $ dr=|d{\bf x}|$,  $\Delta r=r_1-r_0$ and $ n^{\mu} =dx^{\mu}/dr$. The 
integration is taken along the null geodesic connecting two points $x$ and 
$x'$, and  
\be
 G^{R}_{\mu\nu\rho\sigma}(x,x')= \langle \psi| h_{\mu\nu}(x) h_{\rho\sigma}(x')+
 h_{\mu\nu}(x') h_{\rho\sigma}(x)|\psi \rangle 
\ee 
is the graviton Hadamard  function, understood to be suitably renormalized.
The gauge invariance of  $\Delta t$, as given by Eq.~(\ref{eq:MDT}),  
has been analyzed recently~\cite{YUF}. 

In this subsection, we wish to rederive $\Delta t$ using the geodesic deviation 
equation. This derivation allows us to see the gauge invariance more clearly,
and to discuss the issue of Lorentz invariance of lightcone fluctuations.
 Let us consider a pair of  timelike geodesics with tangent vector $u^{\mu}$,  
and $n^{\mu}$ as a unit spacelike 
vector pointing from one geodesic to the other (See Fig. 1). 
\begin{figure}[hbtp]
\begin{center}
\leavevmode\epsfxsize=1.4in\epsfbox{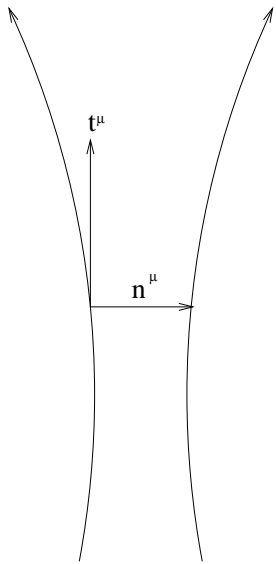}
\end{center}
\caption{ A pair of nearby timelike geodesics. Here  $u^{\mu}$ is a tangent 
vector along the 
geodesic, while $n^{\mu}$ is  a unit spacelike vector pointing from one geodesic to the other.  }
\label{geodesic}
\end{figure}
The geodesic deviation equation is given by
\be
{D^2n^{\mu}\over d\tau^2} = 
-R^{\mu}_{\alpha\nu\beta}u^{\alpha}n^{\nu}u^{\beta}\,,
\ee
where $R^{\mu}_{\alpha\nu\beta}$ is the Riemann tensor.
The relative acceleration per unit proper length of particles on the 
neighboring geodesics is
\be
\alpha \equiv n_{\mu}{D^2n^{\mu}\over d\tau^2}= 
-R_{\mu \alpha\nu\beta}n^{\mu}u^{\alpha}n^{\nu}u^{\beta}\,.
\ee
Thus if $ds$ is the spatial distance between the two particles, then
$\alpha\,ds$ is their relative acceleration. 
It follows that the relative change in displacement of the two particles 
after a proper time $T$ is 
\be
ds\,\int^T_0\,d\tau\,\int^{\tau}_0\,d\tau' \alpha(\tau',0)
\,,
\ee
Now consider the case of two observers (particles) separated by a finite 
initial distance $s_0$ as illustrated in 
Fig.~(\ref{geodesic2}).

\begin{figure}[hbtp]
\begin{center}
\leavevmode\epsfxsize=2.2in\epsfysize=2.0in\epsfbox{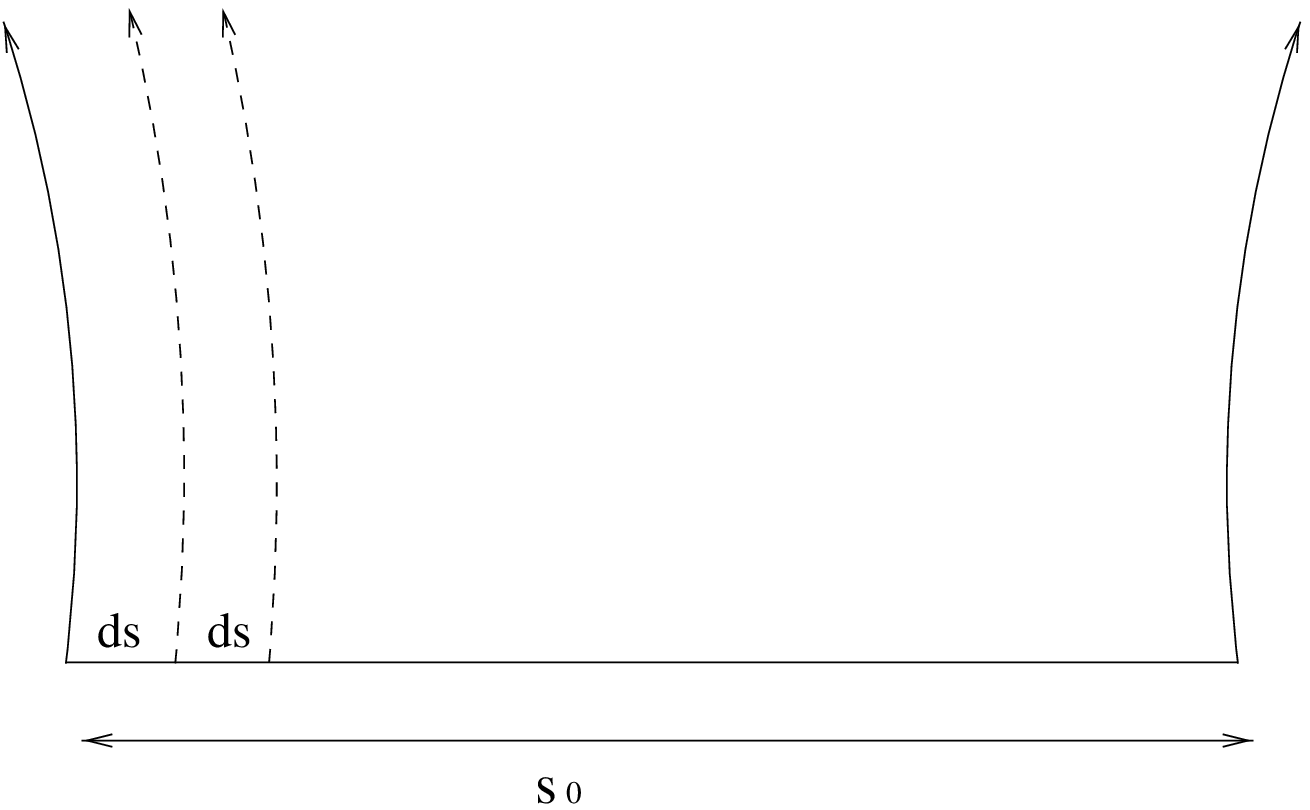}
\end{center}
\caption{ Two timelike geodesics separated by a finite interval containing 
an infinite number of nearby geodesics .}
\label{geodesic2}
\end{figure}
 We can find the relative change in displacement of these two observers by 
integrating on $s$:
\be
\Delta s = \int^{s_0}_0\,ds\,\int^T_0\,d\tau\,\int^{\tau}_0\,d\tau' 
\alpha(\tau',0)
\,.
\ee
This is the relative displacement measured at the same moment of proper time 
for both observers.

\begin{figure}[hbtp]
\begin{center}
\leavevmode\epsfxsize=2.6in\epsfysize=2.6in\epsfbox{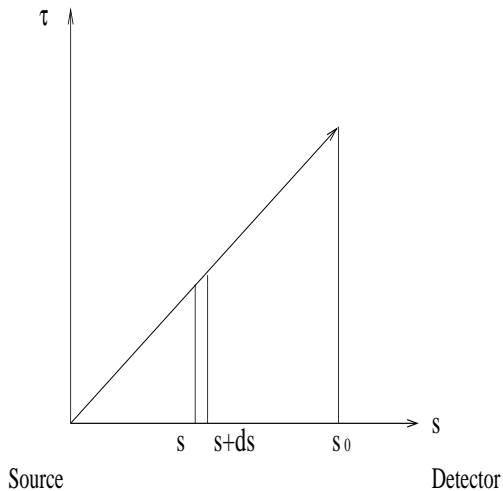}
\end{center}
\caption{The displacement, $\Delta s$, between a source and a detector is
given by an integration within the triangular region. }
\label{fig=geodesic3}
\end{figure}

Let us now consider a light signal sent from one observer to the other. 
If $\alpha=0$, the distance traveled by the 
light ray is $s_0$. When $\alpha\neq 0$, this distance becomes $s_0+\Delta s$, 
where now
\be
\Delta s = \int^{s_0}_0\,ds\,\underbrace{\int^s_0\,d\tau\,\int^{\tau}_0\,
d\tau' \alpha(\tau',s)}
\,.
\ee
Here the  under-braced integral is the displacement per unit $s$ of a pair of 
 observers at a distance $s$ from the source. The domain of the final two
integrations is illustrated in Fig.~(\ref{fig=geodesic3}).

If gravity is quantized, the Riemann tensor will fluctuate around an average 
value of zero due to quantum gravitational
vacuum fluctuations. This leads to $\langle \alpha \rangle=0 $, and hence  
 $\langle \Delta s \rangle=0$. Notice here
that   $ \alpha$ becomes a quantum operator when metric perturbations are 
 quantized.
However, in general,  $\langle (\Delta s )^2\rangle\neq0$,  and we have
\be
 \langle (\Delta s )^2\rangle=
\int^{s_0}_0\,ds_1\int^{s_0}_0\,ds_2\int^{s_1}_0\,d\tau_1\int^{\tau_1}_0\,
d\tau'_1\int^{s_2}_0\,d\tau_2\int^{\tau_2}_0
\,d\tau'_2\,\langle \alpha(\tau'_1,s_1)\alpha(\tau'_2,s_2)\rangle\,.
\label{eq:MDD}
\ee
Thus the root-mean-squared fluctuation in the flight path is 
$\sqrt{\langle (\Delta s )^2\rangle}$, which can also be
understood as a fluctuation in the speed of light. It entails an intrinsic 
quantum uncertainty in the measurement of distance. Therefore, spacetime 
becomes fuzzy at a scale characterized by  
$\sqrt{\langle (\Delta s )^2\rangle}$. 
 The integrand in Eq.~(\ref{eq:MDD}) is obviously invariant under any
 coordinate transformation while the 
integral is gauge invariant within the linear approximation.

We now wish to show that this gauge-invariant quantity is the same as 
Eq.~(\ref{eq:MDT}) when calculated in the transverse-tracefree (TT)
gauge. Choose a coordinate system where the source and the detector are both at 
rest, and suppose that the light 
ray propagates in the $x$-direction, then we have 
\be
u^{\mu}=(1,0,0,0)\,,
\ee
\be
n^{\mu}=(0,1,0,0)\,,
\ee
and 
\be
\alpha =R_{xtxt}=-{1\over2}h_{xx,tt}\,. 
\ee
Substitution of the above results into Eq.~(\ref{eq:MDD}) leads to
 \ben
 \langle (\Delta s )^2\rangle &=&
\int^{r}\,dx_1\int^{r}_0\,dx_2\int^{x_1}_0\,dt_1\int^{t_1}_0\,dt'_1
\int^{x_2}_0\,dt_2\int^{t_2}_0
\,dt'_2\,\langle \alpha(t'_1,s_1)\alpha(t'_2,s_2)\rangle \nonumber\\
&=&{1\over 4}\,\int^r_0\,dx_1\int^r_0\,dx_2\, \langle h_{xx}(x_1,x_1)
h_{xx}(x_2,x_2)\rangle={1\over r}
\langle \sigma_1^2\rangle\,,
\een
where we have set $s_0=r$ and used the fact that along the light ray $x=t$.
Thus, one has
\be
\Delta t= { \sqrt{\langle \sigma_1^2\rangle}\over r}=\sqrt{\langle 
(\Delta s )^2\rangle }
\ee
which also demonstrates the gauge-invariance of $\Delta t$.  

Now we wish to discuss the rather subtle issue of the relation of lightcone
fluctuations to Lorentz symmetry. It is sometimes argued that lightcone
fluctuations are incompatible with Lorentz invariance. The most dramatic
illustration of this arises when a time advance occurs, that is, when a
pulse propagates outside of the classical lightcone. In a Lorentz invariant
theory, there will exist a frame of reference in which the causal order
of emission and detection is inverted, so the pulse is seen to be detected 
before it was emitted. Thus the lightcone fluctuation phenomenon, if it
is to exist at all, seems to be incompatible with strict Lorentz invariance.

Our view of the situation is the following: lightcone fluctuations respect
Lorentz symmetry on the average, but not in individual measurements. The
symmetry on the average insures that the mean lightcone be that of classical
Minkowski spacetime. The average metric is that of Minkowski spacetime
provided that $\langle h_{\mu\nu} \rangle = 0.$ However, a particular pulse
effectively measures a spacetime geometry which is not  Minkowskian and
not Lorentz invariant. A simple model may help to illustrate this point.
Consider a quantum geometry consisting of an ensemble of classical
Schwarzschild spacetimes, but with both positive and negative values for the
mass parameter $M$. (The fact that the $M < 0$ Schwarzschild spacetime has
a naked singularity at $r=0$ need not concern us. For the purpose of this
model, we can confine our discussion to a region where $r \gg |M|.$)
Suppose that this ensemble has $\langle M \rangle= 0$, 
but $\langle M^2 \rangle \not= 0$. 
It is well known that light propagation in a $M > 0$ Schwarzschild spacetime
can exhibit a time delay relative to what would be expected in flat spacetime.
This is the basis for the time delay tests of general relativity using
radar signals sent near the limb of the sun. In the present model, however, the
time difference is equally likely to be a time advance rather than a time delay.
A measurement of the time difference amounts to a measurement of $M$. This model
is Lorentz invariant on the average because $\langle M \rangle= 0$ and the 
average
spacetime is Minkowskian. However, a specific measurement selects a particular
member of the ensemble, which is generally not Lorentz invariant.

In addition to the fact that the mean metric is Minkowskian, there is another
sense in which lightcone fluctuations due to compactification 
exhibit average Lorentz invariance.  
Note  that $\Delta s$, and hence $\Delta t$, depends on the Riemann tensor 
correlation function $\langle R_{xtxt}(x_1)R_{xtxt}(x_2)\rangle$, which is 
invariant under Lorentz boosts along the $x$-axis. Thus if we were to repeat
 the above
calculations of $\Delta s$ in a second frame moving with respect to the first, 
the result will be the same. In both cases one is assuming that the detector
is at rest relative to the source. This is a reflection of the Lorentz 
invariance of the spectrum of fluctuations, which is exhibited by the 
compactified flat spacetimes studied in this paper, but not by the 
 Schwarzschild spacetime with a fluctuating mass.


\subsection{Graviton two-point functions }

 
We shall use a quantization of the linearized gravitational perturbations 
$h_{\mu\nu}$
which retains only physical degrees of freedom. That is,  we are going to work 
in the 
TT gauge in which the gravitational perturbations have only spatial components 
$h_{ij}$, 
 satisfying the transverse,  $\partial^i h_{ij}=0$,  and tracefree, $ h^i_i=0$ 
conditions. Here $i,j$ run from 1 to $3+n=d-1$.  These $2d$ conditions remove all of 
the gauge degrees of freedom and leave ${1\over 2}(d^2-3d)$ physical degrees of 
freedom. We have 
\begin{equation}
h_{ij} = \sum_{{\bf k},\lambda}\, [a_{{\bf k}, \lambda} e_{ij} ({{\bf k}, \lambda})
  f_{\bf k} + H.c. ].
\end{equation}
Here H.c. denotes the Hermitian conjugate, $\lambda$ labels the ${1\over 2}(d^2-3d)$ independent 
 polarization states,   $f_{\bf k}$  is the mode function, 
 and the $e_{\mu\nu} ({{\bf k}, \lambda})$
are polarization tensors. (Units in which $32\pi G_d =1$, where $G_d$ is
Newton's constant in d dimensions  and in which $\hbar =c =1$  will be used in this paper.) 

Let us now calculate the  Hadamard function, $G_{\mu\nu\rho\sigma}(x,x')$, 
 for gravitons in the Minkowski vacuum state in the transverse tracefree gauge. 
It follows that
\begin{equation}
 G_{ijkl}(x,x')=\frac{2 Re}{(2\pi)^{d-1}}\int\,{d^{d-1}{\bf k}
\over{2 \omega}}
\sum_{\lambda}
\, e_{ij} ({{\bf k}, \lambda}) e_{kl} ({{\bf k}, \lambda})
e^{i{\bf k} \cdot({\bf x}-{\bf x'})}e^{-i\omega(t-t')}\,.
\end{equation}
The summation of polarization tensors in the transverse tracefree gauge is
 (See the Appendix
in Ref.~\cite{YUF}.\footnote{ The tensorial argument given there applies in any 
number of dimensions}  )
\begin{eqnarray}
\sum_{\lambda}\,&& e_{ij} ({{\bf k}, \lambda})\, e_{kl} ({{\bf k}, \lambda})=
\delta_{ik}\delta_{jl}
+\delta_{il}\delta_{jk}-\delta_{ij}\delta_{kl}
+\hat k_i\hat k_j \hat k_k\hat k_l\nonumber\\ 
&& +\hat k_i \hat k_j \delta_{kl} +\hat k_k \hat k_l \delta_{ij}-\hat k_i 
\hat k_l \delta_{jk}
-\hat k_i \hat k_k \delta_{jl}-\hat k_j \hat k_l \delta_{ik}-\hat k_j 
\hat k_k \delta_{il}\,,
\end{eqnarray}
where 
$\hat k_i=\frac{ k_i}{ k}\,.$
 We find that 
\ben
G_{ijkl}&=&
2F_{ij}\delta_{kl} +2F_{kl}\delta_{ij}-2F_{ik}\delta_{jl}-2F_{il}
\delta_{jk}-2F_{jl}\delta_{ik} -2F_{jk}\delta_{il}\nonumber\\
&&+2H_{ijkl}
 +2D(x,x')(\delta_{ik}\delta_{jl}+\delta_{il}\delta_{jk}-\delta_{ij}
\delta_{kl})\,.
\een
Here $D(x,x')$,  $F_{ij}(x,x')$ and $H_{ijkl}(x,x')$ are functions which are
 defined as follows:
\be
 	D^n(x,x')=
{Re\over{(2\pi)^{3+n}}}\int\, {d^{3+n}{\bf k}\over{2 \omega}}e^{i{\bf k} \cdot({\bf x}-{\bf x'})}e^{-i\omega(t-t')} \,,
\label{eq:Dfunc}
\ee

\be
 	F^n_{ij}(x,x')={Re\over{(2\pi)^{3+n}}}\partial_i\partial'_j\int\, 
{d^{3+n}{\bf k}\over{2 \omega^3}}e^{i{\bf k} \cdot({\bf x}-{\bf x'})}
e^{-i\omega(t-t')} \,,
\label{eq:Ffunc}
\ee
and 
\be
 	H^n_{ijkl}(x,x')={Re\over{(2\pi)^{3+n}}}
   \partial_i\partial^{\prime}_j\partial_k\partial_l^{\prime}
\int\, {d^{3+n}{\bf k}\over{2 \omega^5}}e^{i{\bf k} \cdot({\bf x}-
{\bf x'})}e^{-i\omega(t-t')}
\,.
\label{eq:Hfunc}
\ee
These functions are calculated in the Appendix. 
Let  
\be
R=|{\bf x}-{\bf x}'|,  \quad\quad \Delta t= t-t'\,.
\ee
 For $n=2m+1$, an odd number of extra dimensions, the results are 
\be
D^{2m+1}=
  \left\{\begin{array}{ll}
            {(2m+1)!!\over 2(2\pi)^{m+2}}{1\over (R^2-\Delta t^2)^{m+{3\over2}}}\,,
                      &\mbox{ for}\, R^2>\Delta t^2\,,\\
             0 &\mbox{ for}\, R^2<\Delta t^2\,,        
   \end{array}\right.
\label{eq:Dodd}
\ee
\be
F^{2m+1}_{ij}=-{1\over 2(2\pi)^{m+2}}\,\partial_i\partial_j'\left[
{(2m)!!\over R^{2m}}S(0)\,\sum_{k=0}^m{(2k+1)!!\over (2k)!!(2k+1)(2k-1)}
{R^{2k}\over (R^2-\Delta t^2)^k}\right]\,,
\label{eq:Fodd}
\ee
and
\ben
H_{ijkl}^{2m+1}&=&{1\over 2(2\pi)^{m+2}}\,\partial_i\partial_j'\partial_k\partial_l'\biggl\{
 {(2m)!!\over R^{2m-2}}\biggl[{1\over2}Q(1)\nonumber\\
&&+\,\sum_{j=0}^{m-2}
{(m-j-1)(2j+1)!!\over 4m(j+1)(2j)!!(2j+1)(2j-1)}
{R^{2j}\over (R^2-\Delta t^2)^j}S(0)\,\Biggr]\,.\nonumber\\
\label{eq:Hodd}
\een
In particular, for $n=1$ ($d=5$), we have
\be
H^1_{ijkl}=0
\ee
Here 
\be
S(0)=\left\{\begin{array}{ll} 
                 {\sqrt{R^2-\Delta t^2}\over R^2} &\mbox{ for}\, R^2>\Delta t^2\,,\\
                 0 &\mbox{ for}\, R^2<\Delta t^2\,,        
	    \end{array}\right.
\ee
and
\be
Q(1)=\biggl({1\over3}-{\Delta t^2\over 3R^2}\biggr)S(0)\,.
\ee

For $n=2m$, an even number of extra dimensions, we have 
\be
D^{2m}={2^m m!\over (2\pi)^{m+2}}{1\over (R^2-\Delta t^2)^{m+1}}\,,
\label{eq:Deven}
\ee
\ben
F^{2m}_{ij}&=&{1\over (2\pi)^{m+2}}\,\partial_i\partial_j'\biggl\{
{(2m-1)!!\over R^{2m}}\biggl[1-{\Delta t\over4 R}
\ln\biggl({R+\Delta t\over R-\Delta t}\biggr)^2\,\nonumber\\
&&
-{1\over R^2}\sum_{k=2}^m{2^{k-2}\Gamma(k-1)\over (2k-1)!!}
{R^{2k}\over (R^2-\Delta t^2)^{k-1}}\biggr]\biggr\}\quad\quad\quad m\geq 1\,,
\nonumber\\
\label{eq:Feven}
\een
and
\ben
H_{ijkl}^{2m}&=&{1\over (2\pi)^{m+2}}\,\partial_i\partial_j'
\partial_k\partial_l'\biggl\{{(2m-1)!!\over R^{2m-4}}\biggl[\,
{\Delta t\over24R^3}\biggl({\Delta t^2\over R^2}-1\biggl)
\ln\biggl({R+\Delta t\over R-\Delta t}\biggr)^2
-{1\over18R^2}\nonumber\\
&&\quad
-{\Delta t^2\over6R^4}
-\sum_{k=3}^m\,{1\over (2k-1)(2k-3)}\left({1\over R^2}-
{\Delta t\over4 R^3}\ln\biggl({R+\Delta t\over R-\Delta t}\biggr)^2\,\right)
\nonumber\\
&& +{1\over R^4}\sum_{j=2}^{m-2}\,
{(m-j-1)2^{j-2}\Gamma(j-1)\over (2m-1)(2j+1)!!}{R^{2j}\over 
(R^2-\Delta t^2)^{j-1}}\,
\biggr]
\biggr\}\quad\quad\quad m\geq 2\,.
\nonumber\\
\label{eq:Heven}
\een
For the case $n=2$ ($d=6$), we have for $H$
\be
H^{2}_{ijkl}={1\over (2\pi)^{3}}\,\partial_i\partial_j'\partial_k\partial_l'\,
\Biggl[
-{1\over6}\ln(R^2-\Delta t^2)
-{\Delta t^2\over6R^2}+{\Delta t\over8R}\biggl({\Delta t^2\over3 R^2}-1\biggl)
\ln\biggl({R+\Delta t\over R-\Delta t}\biggr)^2
\Biggr]\,.\nonumber\\
\ee
The case of $n=0$ ($d=4$)  was given in Ref.~\cite{YUF}. 


\section{Periodic compactification scenario}


Let us now suppose that  the extra $n$ dimensions $z_1,...,z_n$ are compactified with
 periodicity lengths $L_1,...,L_n$, namely spatial points $z_i$ and $z_i+L_i$ are 
identified.  For simplicity, we shall assume in this paper that 
$L_1=...=L_n=L$. The effect of imposition of the periodic boundary conditions 
on the extra dimensions is to restrict the field modes to a discrete set
\be
f_{\bf k} = (2\omega (2\pi)^{3}L^n)^{-{1\over 2}}  
e^{i({\bf k \cdot x} -\omega t)} \, ,
	\label{eq:mode2}
\ee
with
\be
k_{i}={2\pi m_i\over L}, \quad i=1,...,n,\quad m_i=0,\pm 1, \pm 2, \pm 3,...\,.
\ee 
Let us denote the associated vacuum state by $|0_L\rangle$. 
 In order to calculate the 
gravitational vacuum fluctuations due to compactification of  extra dimensions, 
we need the 
renormalized graviton Hadamard function with respect to the vacuum state 
$|0_L\rangle$, 
$G^{R}_{\mu\nu\rho\sigma}(x,x')$, which is given by a multiple
 image sum of the 
corresponding Hadamard function for the Minkowski~\footnote{By Minkowski we 
mean flat spacetime with all
 dimensions uncompactified}  vacuum, $G_{\mu\nu\rho\sigma}$ :
\be
 G_{\mu\nu\rho\sigma}^{ R}(t, z_i,\,t',z_i')
 =\prod_{i=1}^n{\sum_{m_i=-\infty}^{+\infty}}^{\prime}
G_{\mu\nu\rho\sigma}(t, z_i   , \,t',z_i' +m_iL)\, .
\ee
Here the prime on the summation indicates that the $m_i=0$ term is excluded and 
the notation 
\be
(t,\vec{x}, z_1,.., z_{n})\equiv(t, z_{i}) 
\ee
 has been adopted. 
 
We are mainly concerned about how lightcone fluctuations arise in the usual 
uncompactified space as a result of compactification of extra dimensions. So
we shall
examine the case of a light ray propagating in one of the uncompactified
dimensions. Take the direction to be along the $x$-axis in our four dimensional
world, then the relevant graviton
two-point function is $G_{xxxx}$, which can be expressed as
\ben
G_{xxxx}(t,\vec{x},z_i,\,t',\vec{x}',z_i') &=& 
2 \biggl[ D(t,\vec{x},z_i,\,t',\vec{x}',z_i')
 - 2 F_{xx} (t,\vec{x},z_i,\,t',\vec{x}',z_i')\nonumber\\
&& + H_{xxxx}(t,\vec{x},z_i,\,t',\vec{x}',z_i') \biggr]\,.
\label{eq:Gx}
\een
Assuming that the propagation goes from point $(a,0,...,0)$ to point 
$(b,0,...,0)$, 
we have 
\ben
\langle \sigma_1^2 \rangle &=&{1\over 8}(b-a)^2
\int_{a}^{b} dx \int_{a}^{b} dx'\, G_{xxxx}^{R }(t,x,{\bf 0},\,t',x',{\bf 0},),
 \,\nonumber\\
&=& {1\over 8}(b-a)^2
\int_{a}^{b} dx \int_{a}^{b} dx'\,
\prod_{i=1}^n{\sum_{m_i=-\infty}^{+\infty}}^{\prime} 
G_{xxxx}(t,x,{\bf 0},\,t',x',0,0, m_1L,...,m_iL)\,.\nonumber\\
\label{eq:sigma1}
\een
With  these results, we can in principle calculate lightcone fluctuations in 
spacetimes
with arbitrary number of flat extra dimensions.  In what follows, we first 
examine some particular cases, then make some observations for the general case.


\subsection{ The five dimensional Kaluza-Klein model}
\label{sec:5DKK}


One of the most intriguing and elegant ways of unifying gauge field theories 
with gravitation 
is the higher-dimensional generalizations of Kaluza-Klein theory. The original 
suggestion of Kaluza and Klein \cite{ACF} was that electromagnetism and 
general relativity could be unified by 
starting with a five-dimensional version of the latter and then somehow 
arranging for 
the fifth dimension to become unobservable. This idea was further generalized
 to higher 
dimensions in  attempts to unify non-Abelian gauge fields with gravitation and 
has been 
extensively studied in recent years in the context of supergravity and 
superstring theories.
In the course of  investigation of new features arising from the introduction 
of  extra 
dimensions, the five-dimensional Kaluza-Klein theory has always been taken as 
a prototypical
 model to carry out explicit calculations to obtain a basic understanding. 
This is also our strategy here. In this subsection, we will derive the
results which were summarized in Ref.~\cite{YUF2}.

\subsubsection{Calculation of $\Delta t$}

To begin, let us examine the influence of the compactification of the fifth 
(extra) dimension on the light propagation in our four dimensional world, by considering a light 
ray traveling along the $x$-direction from point $a$ to point $b$, which is perpendicular to the 
direction of compactification. 
Define 
\be
\rho=x-x',\quad\quad b-a=r\,
\ee
and note the fact that the integration in Eq.~(\ref{eq:sigma1}) is to be 
carried out along 
the classical 
null geodesic on which $ t-t' =\rho$. Then we obtain, after performing the 
differentiation
in both $D$ and $F_{xx}$, 
\ben
&&G_{xxxx}(t,x,0,0,0, t',x',0,0,mL')\nonumber\\
&& \quad =
{1\over 4\pi^2}\Biggl[-{\rho^6\over(\rho^2+ m^2L ^2)^3 |mL|^3}
-{7\rho^4\over (\rho^2+ m^2L^2)^3 |mL|}\nonumber\\
&&\quad\quad
+{9\rho^2|mL|\over (\rho^2+m^2L^2)^3}
- {|mL|^3\over (\rho^2+mL^2)^3}\Biggr]\,.
\een
Thus, we have
\ben
\langle \sigma_1^2 \rangle &=&{1\over 8}r^2
\int_{a}^{b} dx \int_{a}^{b} dx'\, G_{xxxx}^{R }(t,x,{\bf 0},\,t',x',{\bf 0},),
 \,\nonumber\\
&=& {1\over 8}r^2
\int_{a}^{b} dx \int_{a}^{b} dx'\,{\sum_{m=-\infty}^{+\infty}}^{\prime} 
G_{xxxx}(t,x,{\bf 0},\,t',x',0,0, mL)\nonumber\\
& =&{r^2\over 32\pi^2L}{\sum_{m=1}^{\infty}}\,\left [
{8\over m}\ln\bigl(1+{\gamma^2\over m^2}\bigr)-{2\gamma^2\over m^3}-
{8\gamma^2\over (\gamma^2+m^2)m}\right]\,,
\een
where we have introduced a dimensionless parameter $\gamma=r/L$.  
We are interested here 
in the case in which
$\gamma \gg 1$. It then follows that the summation is dominated, to the 
leading order, by
the second term, 
\be
\langle \sigma_1^2 \rangle_R \approx - 
{r^2\over 16\pi^2L} {\sum_{n=1}^{\infty}}\,{\gamma^2\over n^3}
 = -{\zeta(3) r^2\gamma^2\over 16\pi^2L}\,,
\label{eq:sigmaFix}
\ee
where $\zeta(3)$ is the Riemann-zeta function.
So, the mean deviation from the classical propagation time due to the lightcone 
fluctuations
is
\be
\Delta t \approx 
\sqrt{ {\zeta(3)\over 16\pi^2L}}\,\gamma=\sqrt{ {\zeta(3)\over 16\pi^2L}}\sqrt{32\pi G_5}\,\gamma
=\sqrt{ {2\zeta(3)G_4\over\pi}}\gamma\approx \biggl({r\over L}\biggr)\,t_{pl}\,.
\label{eq:tkk}
\ee
Here we have used the fact that $G_5=G_4L$, and   
$t_{pl} \approx 5.39\times 10^{-44}s$ is the Planck time.

This result reveals that here the mean deviation in the arrival time increases 
linearly\footnote{ In the usual four dimensional case with one
 compactified spatial dimension $\Delta t$ grows linearly with the square root 
of $r$ (see Ref.~\cite{YUF}).} 
 with $r$ 
and  inversely with the size of the extra dimension.
 This effect is counter-intuitive in the sense that it grows as the size of the 
compactified dimension decreases. When
$r$ is of cosmological dimensions and $L$ is sufficiently small, the effect
is potentially observable. 

\subsubsection{Choice of Renormalization and Changes in $L$.}

 Note that here $\langle \sigma_1^2 \rangle_R$ has been renormalized to zero 
as $L\rightarrow \infty$.  This is the
most natural choice of renormalization, corresponding to the effect of the
graviton fluctuations vanishing in the limit of noncompactified spacetime.
This is analogous to setting a Casimir energy density to zero in the limit
of infinite plate separation,  However,  if 
instead of renormalizing $\langle \sigma_1^2 \rangle$ against the vacuum with 
respect to 
$L\rightarrow \infty$, 
we take the manifold with compactified extra dimensions at some fixed sizes 
$L$ to have 
$\langle\sigma_1^2\rangle_R=0$,  then the lightcone 
fluctuations could seem to be renormalized away. The latter renormalization 
scheme is 
 a logical possibility that we can consider,  although it is unnatural as 
there seems to be
 nothing in the theory which picks out a particular
finite value of $L$. In any case, if $L$ is somehow 
 allowed to vary, 
for example, as the universe evolves, then lightcone fluctuations would produce 
noticeable 
effects,  as one could at most set $\langle \sigma_1^2 \rangle_R =0$ at one
 point
along the path of a light ray.  It is particularly so, 
when we try to detect the spread in the arrival times of pulses from distant 
astrophysical 
sources, where we are  looking back in  time. Hence significant lightcone 
fluctuations may arise
no matter what renormalization scheme one chooses if the size $L$ is allowed 
to vary.

To get an understanding for the case of a changing $L$,  let us assume that 
$L$ changes 
with  time at an extremely small rate, which is in fact required by experimental data on the time 
evolution of fundamental constants (see, for example, 
Refs.~\cite{KPW,JDB,MM,AIS}). Then the evolution of $L$ can be reasonably 
well approximated
by a linear function of time:~\footnote{Strictly speaking, the functional 
dependence of $L$ on time 
should be given by a yet-unknown underlying dynamical theory which governs 
how the extra dimensions 
evolve.  However, the assumption of a linear dependence is good enough for 
our purpose of getting 
a basic idea about how the variation of $L$ over time would affect our 
results.}  
\be
L=L_i+\alpha t
\ee
Using this expression for $L$ and redoing the calculations~\footnote{It is worth pointing out here that 
the graviton two-point function  obtained simply by replacing the constant $L$ with a changing one is not
the exact two-point function that satisfies the appropriate equations, but it is a very good leading 
order approximation provided that $\alpha \ll 1$},  one finds,
\ben
\langle \sigma_1^2 \rangle_R &=&{r^2\over 32\pi^2L_f}{\sum_{m=1}^{\infty}}\,\biggl( {L_f\over L_i}\biggr)^2
\,\left [
{8\over m}\ln\bigl(1+{\gamma^2\over m^2}\bigr)-{2\gamma^2\over m^3}-{8\gamma^2\over (\gamma^2+m^2)m}\right]\nonumber\\
&\approx& -{\zeta(3) r^2\gamma^2\over 16\pi^2L_f}\,\biggl( {L_f\over L_i}\biggr)^2\,,
\label{eq:sigmaVari}
\een
where $L_i$ is the initial compactification size when  the light ray 
is emitted, $L_f=L_i+\alpha r$ is the final size at the time of reception  and
 $\gamma= r/L_f$. Here $\langle \sigma_1^2 \rangle_R$ is renormalized with respect to $ 
L\rightarrow \infty$. Therefore, one has for the mean time deviation from the classical 
propagation time 
\be
\Delta t={\sqrt{\langle \sigma_1^2 \rangle_R}\over r}\approx\biggl( {L_f\over L_i}\biggr) \biggl({ r\over L_f}\biggr)t_{pl}\,.
\label{eq:Tvari}
\ee

Another possibility, as we have mentioned earlier,  is to renormalize $\langle \sigma_1^2 \rangle$
 against that corresponding to the current size $L_f$, which implements the idea of setting the 
renormalized  $\langle \sigma_1^2 \rangle_R$ to be zero if $L$ is fixed always or at least during 
the propagation of the light. This is accomplished by 
taking the difference between Eq.~(\ref{eq:sigmaVari}) and Eq.~(\ref{eq:sigmaFix}) with $L$ being replaced by $L_f$ to obtain
 \be
\langle \sigma_1^2 \rangle_R 
\approx \Biggl[1-\biggl( {L_f\over L_i}\biggr)^2\Biggr]{\zeta(3) r^2\gamma^2\over 16\pi^2L_f}
\,,
\ee
which leads to 
\be
\Delta t= {\sqrt{\langle \sigma_1^2 \rangle_R}\over r}\approx\sqrt{\biggl|1-\biggl( {L_f\over L_i}\biggr)^2\biggr| }\biggl({ r\over L_f}\biggr)t_{pl}\,.
\label{eq:Tvari1}
\ee
Equations~(\ref{eq:Tvari}) and ~(\ref{eq:Tvari1}) demonstrate clearly
 that no matter what renormalization 
scheme is employed, one gets a nonzero lightcone fluctuation effect as long as $L$ is changing.  
We want to point out here again that renormalizing 
$\langle \sigma_1^2 \rangle$ with respect to $L\rightarrow \infty$ is far
 more natural than to
 a particular finite size $L_f$, since the latter seems to pick out a preferred 
size $L_f$ without any 
convincing reason to do so.

\subsubsection{Correlation of Pulses}
\label{sec:corr}
 
The fluctuation in the flight time of pulses, $\Delta t$, can apply to the 
successive wave crests of a plane wave. This leads to a broadening of spectral lines from a distant
 source. Note, however, that $\Delta t$ is the expected variation in the arrival times of two 
successive crests only when the successive pulses are uncorrelated \cite{Ford96}.
 To determine the 
correlation, we need to  compare $|\langle \sigma_1^2\rangle| $ and 
$|\langle \sigma_1\sigma'_1\rangle| $.  The latter quantity is defined by 
\begin{equation}
\langle \sigma_1 \sigma'_1 \rangle =  
{1\over 8}(\Delta r)^2
\int_{r_0}^{r_1} dr \int_{r_0}^{r_1} dr'
\:\,  n^{\mu} n^{\nu} n^{\rho} n^{\sigma}
\:\, G^{R}_{\mu\nu\rho\sigma}(x,x') \, ,
 \label{eq:sigma11'ex}
\end{equation}
where the $r$-integration is taken along the mean path of the first pulse,
and the $r'$-integration is taken along that of the second pulse. Here
we will assume that $\Delta t \ll \Delta r$, so the slopes, $v$ and $v'$, 
of the two mean paths are 
approximately unity. Thus the two-point function in Eq.~(\ref{eq:sigma11'ex})
will be assumed to be evaluated at $\rho = |{\bf x} -{\bf x'}| = |r -r'|$ 
and $\tau = |t - t'| = |r -r' -t_0|$. If $|\langle \sigma_1\sigma'_1\rangle| \ll |\langle \sigma_1^2\rangle| $, two pulses are uncorrelated, and otherwise they are correlated.

 Let us now suppose the time separation of two pulses is $T$, and  note that 
the relevant graviton two-point function can be expressed as 
\ben
&&G_{xxxx}(t,x,0,0,0, t', x',0,0, nL)|_{t-t'=\rho-T}\nonumber\\
&&\quad\quad={1\over 4\pi^2}\biggl[ 
-{1\over \beta^3}-{8\rho^2\over\beta(\rho^2+n^2L^2)^2}
+{2\over \beta(\rho^2+n^2L^2)}
\nonumber\\
&&\quad\quad\quad\quad+{16\beta \rho^2\over(\rho^2+n^2L^2)^3}
-{4\beta\over(\rho^2+n^2L^2)^2}
+{2n^2L^2\over\beta^3(\rho^2+n^2L^2)}\biggr]\nonumber\\
&&\quad\quad\equiv -{1\over 4\pi^2}{1\over \beta^3}+ {1\over 4\pi^2}f(\rho,n)\,,\nonumber\\
\een
where,
\be
\beta(n,\rho)=(n^2L^2+2\rho T-T^2)^{1/2}\,.
\ee
Utilizing the following integration relation
\be
\int_a^b\,dx\int_a^b\,dx' f(x-x')=
\int_0^r\, (r-\rho)[f(\rho)+f(-\rho)]\,d\rho\,,
\ee
 one finds that
\be
\langle \sigma_1\sigma'_1\rangle=A+B\,,
\ee
where
\ben
A &=&-{r^2\over 16\pi^2}\sum^{\infty}_{n=1}\,\int^r_0\, d\rho\, (r-\rho)
\biggl[{1\over \beta(\rho,n)^3}+{1\over\beta(-\rho,n)^3}\biggr]\nonumber\\
&=&-{r^2\over 16\pi^2L^3 }\sum^{\infty}_{n=1}\,{L^4\over T^2}\biggl( 2\sqrt{n^2-{T^2\over L^2}}
-\sqrt{n^2+{2rT-T^2\over L^2}}-\sqrt{n^2-{2rT+T^2\over L^2}}\biggr)\,,
\nonumber\\
\label{eq:A}
\een
and 
\ben
B &=&{r^2\over 16\pi^2}\sum^{\infty}_{n=1}\,\int^r_0\, d\rho\, (r-\rho)
[f(\rho,n)+f(-\rho,n)]\nonumber\\
&=&{r^2\over 16\pi^2L^3 }\sum^{\infty}_{n=1}\,2\Biggl[
-{2\sqrt{n^2L^2-T^2}\over n^2L^2}+{\beta(r,n)+\beta(-r,n) \over (n^2L^2+r^2)}
+{2\over nL}\ln\biggl({ nL+\sqrt{n^2L^2-T^2}\over{nL-\sqrt{n^2L^2-T^2}}} \biggr]
\nonumber\\
&&\quad - {1\over nL}\ln\biggl({ n^2L^2+rT+nL\beta(r,n) \over{n^2L^2+rT-nL\beta(r,n)}}\biggr)
-{1\over nL} \ln\biggl({ n^2L^2-rT+nL\beta(-r,n) \over{n^2L^2-rT-nL\beta(-r,n)}}\biggr)
\Biggr)\,.\nonumber\\
\een

A few things are to be noticed here: (1) We need to drop those terms in $A$ 
and $B$ when the square root is 
imaginary. (2) It can be shown that the above expression for 
$\langle \sigma_1\sigma'_1\rangle$ reduces to  $\langle \sigma_1^2\rangle $
 when $T=0$, as it should. 
(3) The asymptotic behaviors of the summands when $n\rightarrow \infty$,  are  
 $\sim{ 2\over n^3}$ for $ A$ and $\sim {2\over n^5}$ for $B$, hence both $A$ and $B$ converge.
(4) If $r\gg T, L$, then $A$ dominates over $B$, since the leading order of 
the summand in $A$ is
$\sqrt{r}$ while that in $B$ is a constant independent of $r$ as 
$r\rightarrow \infty$.  Thus  $\langle \sigma_1\sigma'_1\rangle\approx A$.

To proceed, 
let us now assume that $r\gg T$ and $r\gg L$, then 
\be
p\equiv \sqrt{{2rT+T^2\over L^2}}\approx\sqrt{{2rT\over L^2}} \gg 1
\label{eq:p}
\ee
is a huge number. Thus for the third term in Eq.~(\ref{eq:A}), the sum  should only start from $n=p$. We can 
now split the summation into two parts, i.e. terms with $n\leq p$ and those  
with $n>p$. 
Using the asymptotic form of the summand for the part with $n>p$ and defining 
$m=[T/L]$, where $[\;]$ denotes the integer part, one has 
\be
\langle \sigma_1\sigma'_1\rangle \approx -{r^2\over 16\pi^2L^3 }\biggl(
{2L^4\over T^2}\sum^p_{n=m}\sqrt{n^2-m^2}
-{L^4\over T^2}\sum^p_{n=1}\sqrt{n^2+{2rT-T^2\over L^2}}+\sum^{\infty}_p{2r^2\over n^3}\,. 
\biggr)
\ee
Hence, it follows that
 \be
|\langle \sigma_1\sigma'_1\rangle| < {r^2\over 16\pi^2L^3 }\biggl(
{2L^4\over T^2}\sum^p_{n=1}n
+{L^4\over T^2}\sum^p_{n=1}\sqrt{n^2+{2rT-T^2\over L^2}}+\sum^{\infty}_p{2r^2\over n^3}\,. 
\biggr)
\ee
 Let us now  evaluate the above expression term by term. One has, keeping in
mind that $p\gg 1$, that
\ben
&&\sum^p_{n=1}\sqrt{n^2+{2rT-T^2\over L^2}}\lprox  \sum^p_{n=1}\sqrt{ n^2+p^2}
=p\sum^p_{n=1}\sqrt{ n^2/p^2+1}\nonumber\\
&&\quad\quad\approx p^2\int^1_{1/p}\,\sqrt{x^2+1}\,dx
\approx{1\over 2}[\sqrt{2}+\ln(\sqrt{2}+1)]\,p^2 \,,
\een
and 
\be
\sum^{\infty}_p{r^2\over n^3}=-{r^2\over 2}\Psi(2,p)\sim {r^2\over 2}{1\over p^2}
={1\over 4}{L\over r}{L\over T}\,r^2\,.
\ee
Here we have used Eq.~(\ref{eq:p}) and the asymptotic expansion for $\Psi(2,x)$
\be
\Psi(2,x)\approx -{1\over x^2}-{1\over x^3}-{1\over 2x^4}+O(1/x^6)\,,
\ee
 where function $\Psi(n,x)$ is defined as  
\be 
\Psi(n,x)={d^n \psi(x)\over dx^n}\,,
\quad\quad \psi(x)={d\over dx}\ln\Gamma(x)\,,
\ee
Noting that for $p\gg 1$, 
\be
\sum^p_{n=1}n\sim {1\over 2}p^2\,,
\ee
and letting
\be
A=1+{1\over 2}[\sqrt{2}+\ln(\sqrt{2}+1)]\,,
\ee
we finally find
\be
|\langle \sigma_1\sigma'_1\rangle| < {r^2\over 16\pi^2L^3 }\biggl(
A{L^4\over T^2}p^2+{r^2\over 2}{1\over p^2}\biggr)= {r^2\over 16\pi^2L^3 }
\biggl(2A+{1\over2}\biggr){L\over r}{L\over T}\,r^2\,.
\ee

Compare this result with 
\be
|\langle \sigma_1^2 \rangle_R| \approx{ r^2\over 16\pi^2L^3} \zeta(3)\, r^2\,.
\ee
It is  seen that 
 two successive pulses separated by $T$ in time are uncorrelated
($|\langle \sigma_1\sigma'_1\rangle|\ll|\langle \sigma_1^2 \rangle_R|$)
provided that
\be 
r\gg {L^2\over T}\,,
\ee 
or equivalently
\be
T\gg {L^2\over r}={L\over r}\,L\, . 
\label{eq:Leff}
\ee 
However, if $r\ll L$, one can show, by series expanding both $A$ and $B$,  that 
\be
|\langle \sigma_1\sigma'_1\rangle| \approx {r^2\over 16\pi^2L^3 }
\sum^{\infty}_{n=1}{1\over (n^2-{T^2\over L^2})^{3/2}} \,.
\ee
Clearly, in this case,  $|\langle \sigma_1\sigma'_1\rangle|\ll|\langle \sigma_1^2 \rangle_R|$,
if $T\gg L$, and $|\langle \sigma_1\sigma'_1\rangle|\approx |\langle \sigma_1^2 \rangle_R|$, 
when $T<L$.

A few comments are now in order about the physical picture behind our 
correlation results. 
It is natural to expect from the configuration that the dominant contributions 
to the light 
cone fluctuation come from the graviton modes with wavelengths of the order of 
$\sim L$. In other words, the lightcone fluctuates on a typical time scale of
 $\sim 1/L$. If the travel 
distance, $r$, is less than $L$, successive pulses are uncorrelated only when 
their time 
separation is  greater than the typical fluctuation time scale. Otherwise they 
are correlated 
because the quantum gravitational vacuum fluctuations are not significant 
enough in the 
interval between the pulses. However, if $r\gg L^2/T$, then successive pulses 
are in general uncorrelated. Thus the correlation time for large $r$ is of
order $L^2/r$, which is much smaller than the compactification scale $L$.
We can understand this result as arising from the loss of correlation as
the pulses propagate over an increasing distance.

\subsubsection{Observational Limits} 

Suppose that the experimental fractional resolution for a particular 
spectral line of period $T$ is 
$\Gamma$. Then we must have
\be
{\Delta t\over T}\leq \Gamma \, ,
\label{eq:resolution}
\ee
which, with Eq.~(\ref{eq:tkk}), leads to a bound on $L$ of
\be
L\geq {r\,t_{pl}\over \Gamma T}\,,
\label{eq:Lbound}
\ee
assuming   $L$ does not change over time. 
However,  this bound can be trusted only 
when two successive wave crests are uncorrelated, when $\Delta t$ is 
the expected variation in their arrival times. 
The most conservative constraint from this requirement is that $L$
is  smaller than the wavelength of the spectral line, $T$.  Namely, 
\be
{r\,t_{pl}\over \Gamma T}\leq T\,,
\ee
yielding a restriction on the  range of spectral lines that we should use to 
get the lower bound 
\be
T\geq\sqrt{ {r\,t_{pl}\over \Gamma }}\,.
\label{eq:Tbound}
\ee
If Eq.~(\ref{eq:Tbound}) is approximately an equality, then  
Eq.~(\ref{eq:Lbound}) becomes 
\be
L\geq \left( {r  t_{pl}\over  \Gamma} \right)^{{1\over2}}=\left( {r l_{pl}\over \Gamma} \right)^{{1\over2}}\,.
\ee
 Obviously, the optimal lower  bound would  be deduced from the spectral  
lines 
of  distant galaxies, possibly of cosmological distance, with the highest 
observed spectral 
resolution. For astrophysical sources of cosmological distance, spectral 
lines which satisfy 
Eq.~(\ref{eq:Tbound}) will have wavelengths $\gprox 1$mm assuming a resolution 
of $\Gamma\approx 10^{-3}$ and  a cosmological travel distance.
The detection of CO(1$\rightarrow$0) line emission at 2.6 mm from luminous 
infrared galaxies 
and 
quasars\cite{SANDERS} provides the type of data needed to get  a bound. 
According to 
Ref.~\cite{SANDERS}, the observed CO line  resolution for the infrared quasar 
IRAS 07598+6508 
, which is at a distance of 596 Mpc assuming $H_0=75$ km s$^{-1}$ Mpc$^{-1}$, 
is about $\Gamma\approx 10^{-3}$, leading to 
\be
L\gprox  10^{-1}\,{\rm mm}\,.
\ee
Here, the size of the extra dimension has to be macroscopically
large in order not to contradict the astrophysical observation.   
The lower bound given here is within the sensitivity of the recently proposed experiments 
for possible deviations from Newtonian gravity\cite{LCP}.

  Note, however, this bound could be pushed to an 
even higher value if we do not require $L<T$. This is in fact legitimate,  
since for astrophysical sources one usually has both $r\gg L$ and $r\gg T$ 
satisfied, and the
condition for two successive pulses to be uncorrelated is really
\be
r>{L^2\over T}\,.
\label{eq:corrcond}
\ee 
Hence, we can use Eq.~(\ref{eq:Lbound}) to get a bound from experimental data 
as long as the resulting bound, $L$, obeys Eq.~(\ref{eq:corrcond}),  
\be 
L<\sqrt{rT}\equiv L_0\,.
\label{eq:Lcorr}
\ee
Thus, a stronger bound can be
 achieved,  if we can find astrophysical sources of cosmological distance with observed 
spectra of much  smaller wavelengths provided this condition is satisfied.  The observation of $\gamma$ rays from astrophysical sources, such as
gamma-ray bursters (GRBs)~\cite{TP99,MR97},
   provides such an opportunity. The use of these sources  
as probes of possible quantum gravity effects has been explored by a number of 
authors~\cite{ACEMNS,BES,SDB,PK,EFMMN} recently.
Below we will select some of these $\gamma$ ray sources to calculate both the 
lower bound 
$L_b$ from Eq.~(\ref{eq:Lbound}) and $L_0$ from Eq.~(\ref{eq:Lcorr}).  To be 
conservative, we shall assume a resolution of the order of unity for  all the  
gamma rays we are going to consider, since we should at least have this 
resolution before we 
can talk with confidence about the observed energies (or frequencies) of the 
gamma rays. 

In the following Table, we list the source names,  the observed frequencies $\nu$, 
the source distances $D$, and the calculated $L_b$ and $L_0$.   

\begin{center}
{\bf Table: Bounds on L from GRB sources}
\end{center}
\begin{center}
\begin{tabular}{|c||c|c|c|c|}   \hline
Source & D(Mpc) & $\nu(Hz)$ & $L_b(mm)$ & $L_0(mm)$
\\ \hline
GRB 930229~\cite{PNB} & 791  & $4.8\times 10^{19}$ & $6.1\times 10^{4}$  &
$1.2\times 10^{10}$   
\\ \hline
GRB 940217~\cite{KH} & 385  & $4.3\times 10^{24}$ & $2.6\times 10^{9}$  &
$2.8\times 10^{7}$   
\\ \hline
GRB 930131~\cite{MS} & 260  & $1.1\times 10^{23}$ & $4.6\times 10^{7}$  &
$1.5\times 10^{8}$   
\\ \hline
Mrk 421~\cite{SDB} & 112  & $4.8\times 10^{26}$ & $8.7\times 10^{10}$  &
$1.4\times 10^{6}$   
\\ \hline
GRB 980703~\cite{BES} & 1592  & $1.2\times 10^{20}$ & $3.1\times 10^{5}$  &
$1.0\times 10^{10}$   
\\ \hline
GRB 980425~\cite{GRB980425} & 40  & $5.4\times 10^{20}$ & $3.4\times 10^{4}$  & $8.3\times 10^{8}$
\\ \hline
GRB 990123~\cite{GRB990123} & 2400  & $3.0\times 10^{20}$  & $1.2\times 10^{6}$ & $8.5 \times 10^{9}$ 
\\ \hline
\end{tabular} 
\end{center} 

\vspace{0.5cm}
  
From this Table, one can find that the largest lower bound for $L$ comes from GRB930131 and GRB990123, which is 
\be
L\gprox 10^7 {\rm mm}\,.
\ee
For some sources,  Mrk 421,  for instance, one seems to get a much larger bound, which however can 
not be trusted, because the correlation condition $L_b<L_0$ is violated. The physical 
reason is that the frequencies for the gamma rays are so high that even travel over
 a cosmological distance does not wash out the correlation between successive wave crests.
 
In principle,  the above results only apply to the case where $L$ is fixed. When $L$ changes as 
the universe evolves, one should use Eq.~(\ref{eq:Tvari}), or  Eq.~(\ref{eq:Tvari1}),  for the mean 
time deviation, 
$\Delta t$.  In this case, we can  set either a bound on $L$ or
 a bound on the rate of change of $L$,  if we can constrain either the
rate of change of the size of extra dimensions over time or the present size $L$ from
 other 
considerations.  Let $L_f$ be the size of the extra dimension at the present time and 
$L_i$ be  that at the time of  primordial nucleosynthesis, and write   
\be
{L_f\over L_i}=1+\delta\,.
\label{eq:Lrate}
\ee
According to Refs.~\cite{KPW,JDB}, the observational limits,  obtained by 
investigating the effects on the primordial nucleosynthesis of $^4{\rm He}$ 
as a consequence of  the time variation of fundamental constants such as the 
electroweak, strong, and gravitational coupling constants, imply that  
$\delta\lprox 0.01$. 
However,  stronger limits  on $\delta $, which may be less 
reliable,  arise from a detailed study of the 
events which took place $1.8\times 10^9$yr ago on the current site of an open-pit
uranium mine at Oklo in the West African Republic of Gabon~\cite{MM}.  This site gave rise to a 
natural nuclear reactor when it went critical for a period about $1.8\times 10^9$yr 
ago. The Oklo samples constrain the rate of change of extra spatial dimensions  to
 satisfy the limits~\cite{AIS,JDB}
\be
\biggl|{{\dot L}\over L}\biggr|\leq 1.9\times 10^{-19}{\rm yr}^{-1}\,,
\ee
which translates to $\delta\lprox 10^{-10}$. In either case, $\delta\ll 1$. Thus all the results 
obtained so far for a fixed $L$ hold for a changing $L$ if  renormalization with respect to 
$L\rightarrow \infty$ is adopted (refer to Eq.~(\ref{eq:Tvari})). However if one chooses the 
renormalization with respect to the current size $L_f$, 
then combination of  Eq.~(\ref{eq:resolution}) and Eq.~(\ref{eq:Tvari1})
gives rise to 
\be
L\geq \sqrt{\delta ^2+2\delta }\biggl( {rt_{pl}\over T\Gamma}\biggr)
\approx \sqrt{\delta} \biggl( {rt_{pl}\over T\Gamma}\biggr)\,.
\label{eq:Lvaribound}
\ee
As a result, the following considerably 
smaller lower bounds can be deduced from Eq.~(\ref{eq:Lvaribound}) using the CO line data
\ben
L\gprox  10^{-2}\,{\rm mm}\,,&& \quad {\rm for } \quad \delta = 0.01\,,\nonumber\\
L\gprox  10^{-6}\,{\rm mm}\,,&& \quad {\rm for } \quad \delta = 10^{-10}\,, 
\een
and
\ben
L\gprox  10^{6}\,{\rm mm}\,,&& \quad {\rm for } \quad \delta = 0.01\,,\nonumber\\
L\gprox  10^{2}\,{\rm mm}\,,&& \quad {\rm for } \quad \delta = 10^{-10}\,, 
\een
using GRB data.

On the other hand, one can also use our result to place a bound on the rate of change 
of $L$ if the present size of $L$ can be fixed by other considerations. In this 
respect, it is  well-known that the five-dimensional Kaluza-Klein theory 
 provides an explanation of  the quantization of 
electric charge, in the sense that all electric charges are multiples of the
 elementary charge 
\be
e={4\sqrt{\pi G}\over L}\,.
\ee 
The corresponding fine structure constant is then
\be
\alpha_f={4G\over L^2}\,.
\ee
Setting $\alpha_f$ to its present value, $1/137$, we get an estimate of the size 
of the fifth dimension, $L\sim 10^{-31}$ cm, in the original Kaluza-Klein model.  For the case of 
renormalization with respect to $ L \rightarrow \infty$, we can see from 
Eq.~(\ref{eq:resolution}), 
Eq.~(\ref{eq:Tvari}) and Eq.~(\ref{eq:Lrate}) that  we must have $\delta< 1$, 
which is much weaker 
than the existing bounds on the change rate from primordial nucleosynthesis and 
the Oklo samples.  To discuss the case 
of renormalization with respect to $L_f$, the current size,  let us write 
Eq.~(\ref{eq:Lvaribound}) 
 as
\be
\delta\leq \biggl({LT\Gamma\over r t_{pl}}\biggr)^2\,.
\ee
Thus, 
 using $L\sim 10^{-31}$ cm and the same CO line data as before, we find the following 
limit for  change of the extra dimension in the Kaluza-Klein model
\be
\delta\leq 10^{-58}\,,
\ee
or the rate of change by dividing $\delta $ by the travel time, which is $\sim 10^9 $ yr
\be
\biggl|{{\dot L}\over L}\biggr|\leq 10^{-67}{\rm yr}^{-1}\,.
\ee
This is much stronger even than the strongest bound arising from the observational 
limits on the time evolution of fundamental constants.
  
To conclude, we have demonstrated, in the case of one extra dimension, that
the large quantum lightcone fluctuations due to the compactification of the 
extra dimension require either the size of the extra dimension to be macroscopically 
large or rate of change of the extra dimension to be extremely small.  This result seems to rule out the five dimensional Kaluza-Klein 
 theory, or at the very least, place strong limits on the rate of change of the
extra dimension.
We must point 
out that the rate of growth of $\Delta t$ with $r$  depends crucially on the number of 
spatial dimensions. In four dimensions, $\Delta t\propto \sqrt{r}$, while in  five dimensions 
$\Delta t \propto r$.  One expects that in larger number of dimensions, there will be an effect of compactification, but its details need to be determined by explicit calculations for particular models. This is the topic for the next section.


\subsection{ Higher dimensional models}

There was no real reason to extend the Kaluza-Klein idea beyond five dimensions 
until the 
emergence of non-abelian gauge field theories which have had a profound impact
 on theoretical
physics since their invention by Yang and Mills in 1954. It was suggested by 
DeWitt~\cite{DeWitt63} as 
early as in 1963 that a unification of Yang-Mills theories and gravitation 
could be achieved 
in a higher dimensional Kaluza-Klein framework. Nowadays, the possibility of 
unifying all the 
known interactions in nature in higher dimensional spacetimes has been 
actively pursued in the
context of eleven dimensional supergravity and ten dimensional superstring 
theories. A 
necessary ingredient  of all these higher dimensional models is the 
compactification of the
 extra dimensions to a very small size so as to leave the ordinary 
four-dimensional ``large'' 
world.

In this section, 
we examine, case by case, the lightcone fluctuations arising from quantum 
gravitational 
vacuum fluctuations induced by the periodic compactification of extra, flat
spatial dimensions in
 higher dimensional models   up through 11 dimensions  and 
make a conjecture about arbitrary dimensions.


\paragraph{The case with $n=2$} 

 
This is the 6 dimensional spacetime with 2 extra dimensions.  We obtain, 
after performing the 
differentiation in both $F_{xx}$ and $H_{xxxx}$ (See Eqs~(\ref{eq:Feven})
and (\ref{eq:Heven})).
\be
 G_{xxxx}(t,x,{\bf 0},\,t',x',0,0, m_1L, m_2L)\equiv A_2(\rho)+B_2(\rho)\,
\label{eq:G2}
\ee
with 
\be
A_2(\rho)={1\over 4\pi^3}\left[
-{54\rho^6\over (\rho^2+\alpha_2^2)^5}+{102\alpha_2^2\rho^4\over (\rho^2+\alpha_2^2)^5}
-{3\alpha_2^4\rho^2\over 2(\rho^2+\alpha_2^2)^5}
\right]\,,
\ee 
and
\ben
B_2(\rho)&=&{1\over 4\pi^3} \,
\biggl[{6\rho^7\over (\rho^2+\alpha_2^2)^{11\over2}}
-{27\alpha_2^2\rho^5\over (\rho^2+\alpha_2^2)^{11\over2}}
+{27\alpha_2^4\rho^3\over 4(\rho^2+\alpha_2^2)^{11\over2}}
+{3\alpha_2^6\rho\over 8(\rho^2+\alpha_2^2)^{11\over2}}
\biggr]\times\,\nonumber\\
&&\quad \ln\left( \frac {\sqrt{\rho^{2} + \alpha_2^2} + \rho}{\sqrt{\rho^{2} + \alpha_2^{2}} - \rho}\right )^2\,,
\een
where we have introduced a parameter
\be
\alpha_n^2=L^2\sum_{i=1}^n\,m_i^2\,.
\ee
Upon noting that $G_{xxxx}$ is an even function of $\rho$, it follows that,
\be
\int_{a}^{b} dx \int_{a}^{b} dx'\, G_{xxxx}^{R }(t,x,{\bf 0},\,t',x',{\bf 0})=
\prod_{i=1}^2\sum_{m_i=-\infty}^{+\infty}\int_0^r 2(r-\rho)(A_2+B_2) d\rho \,.
\label{eq:IntRelation}
\ee
Performing the integration (integrate by parts for those terms involving 
logarithmic functions), we find
\ben
&&\prod_{i=1}^2\sum_{m_i=-\infty}^{+\infty}\int_0^r 2(r-\rho)(A_2+B_2) 
d\rho \nonumber\\
&&\quad ={1\over2\pi^3L^2 }{\sum_{m_1=1}^{\infty}}^{\prime}
{\sum_{m_2=1}^{\infty}}^{\prime}\Bigg[
\frac{\gamma^2(m_1^2+m_2^2)}{(\gamma^2+m_1^2+m_2^2)^3}
+\frac{10\gamma^4}{3(\gamma^2+m_1^2+m_2^2)^3}\nonumber\\
&&\quad\quad \quad\quad-{8\gamma^6\over 3(m_1^2+m_2^2)(\gamma^2+m_1^2+m_2^2)^3}
-\frac{8\gamma^5+
4(m_1+m_2^2)\gamma^3+(m_1^2+m_2^2)^2\gamma}{2(m_1^2+m_2^2+\gamma^2)^{7/2}}
\nonumber\\
&&\quad\quad \quad\quad\quad\quad
\times\ln\left({\sqrt{m_1^2+m_2^2+\gamma^2}+ \gamma}\over{\sqrt{m_1^2+m_2^2+
\gamma^2}- \gamma} \right)\Bigg]\,,\nonumber\\
	\label{eq:series2}
\een
where $\gamma=r/L$ is a dimensionless parameter. The above double summation 
is by no means easy to evaluate. 
However,  we are  interested in  the case in which the travel distance $r$ is 
much
greater than the size of the extra dimensions $L$, i.e., 
$\gamma \gg 1$. It then follows that the summation is dominated, to the 
leading order, by
\be
-{1\over2\pi^3L^2 }{\sum_{m_1=1}^{\infty}}^{\prime}
{\sum_{m_2=1}^{\infty}}^{\prime}{8\gamma^6\over 3(m_1^2+m_2^2)
(\gamma^2+m_1^2+m_2^2)^3}\,,
\ee
which can be approximated by the following integral when $\gamma \gg 1$
\be
-{4\over3\pi^3L^2 }\int_{1/\gamma}^{\infty}\,dx_1\,\int_{1/\gamma}^{\infty}\,dx_2\,
{1\over (x_1^2+x_2^2)(x_1^2+x_2^2+1)^3}\approx {4\over3\pi^3L^2 }\ln(\gamma)\,, \quad \quad
{\rm as}\,\,\gamma\rightarrow \infty.
\ee
An easy way to see the above behavior is to note that the contribution to the integral
is dominated by the region around $(1/\gamma, 1/\gamma)$ since the integrand dies 
away very quickly as $x_1$ and $x_2$ increase,
and then change to polar coordinates to evaluate the integral while using the fact that  around $(1/\gamma, 1/\gamma)$ the integrand is approximated 
by $1/(x_1^2+x_2^2)$. Therefore one finds that
\be
\langle \sigma_1^2 \rangle_R \approx {r^2\over6\pi^3L^2 }\ln(\gamma)\,,
\ee
and  the mean deviation from the classical propagation time due to the lightcone fluctuations
\be
\Delta t={\sqrt{ \langle \sigma_1^2 \rangle_R }\over r}\approx
\sqrt{ {32\pi G_{6}\over 6\pi^3L^2}}\,\sqrt{\ln(\gamma)}
=\sqrt{ {16G_4\over3\pi^2}}\sqrt{\ln(\gamma)}\approx {4t_{pl}\over \sqrt{3}\pi}\,
\sqrt{\ln(r/L)}\,.
\label{eq:t2}
\ee
Here we have used the fact that $G_{4+n}=G_4L^n$. Note that here the 
$\Delta t $ dependence on 
$\gamma$ is square root of a logarithmic function  which is quite different 
from linear dependence in the case with $n=5$ discussed above.
 So, the mean time deviation here grows much more slowly as the travel distance $r$ increases or 
the compactification scale decreases. As we shall see later, this seems to a general
 feature for $n>5$. 


\paragraph{The case with n=3}


This is the 7 dimensional spacetime with 3 extra dimensions.  
The relevant two-point function can be shown to be
\ben
 &&G_{xxxx}(t,x,{\bf 0},\,t',x',0,0,m_1L,...,m_3L)\nonumber\\
&&={1\over 8\pi^3}\left[
{24\rho^8\over\alpha_3(\rho^2+\alpha_3^2)^6}
-{336\alpha_3\rho^6\over (\rho^2+\alpha_3^2)^6}
+{280\alpha_3^3\rho^4\over 2(\rho^2+\alpha_2^3)^6}
\right]\,,
\label{eq:G3}
\een
which leads to 
\ben
&&\int_{a}^{b} dx \int_{a}^{b} dx'\, G_{xxxx}^{R }(t,x,{\bf 0},\,t',x',{\bf 0})
\nonumber\\
&&\quad=-{1\over4\pi^3L^3 }\prod_{i=1}^3{\sum_{m_1=1}^{\infty}}^{\prime}\biggl(
 {8\gamma^8\over 3(\sum_{i=1}^3m_i^2)^{3\over2}(\gamma^2+\sum_{i=1}^3m_i^2 )^4}\nonumber\\
&&\quad\quad\quad
+{56\gamma^6\over 3(\sum_{i=1}^3m_i^2 )^{1\over2}(\gamma^2+\sum_{i=1}^3m_i^2 )^4}
\biggr)\,.\nonumber\\
\een
 The triple summation is dominated, to the leading order, by the first term when 
$\gamma \gg 1$, which
 again can be approximated by integration to be
\be
-{2\over3\pi^3L^3 }\int_{1/\gamma}^{\infty}\,dx_1\,\int_{1/\gamma}^{\infty}\,dx_2\,
\int_{1/\gamma}^{\infty}\,dx_3
{1\over (\sum_{i=1}^3 x_i^2)^{3\over2}(\sum_{i=1}^3+1)^4}\approx {2\over3\pi^3L^3 }\ln(\gamma)\,, 
\ee
 as $\gamma\rightarrow \infty$.
Thus the mean deviation from the classical propagation time due to the lightcone fluctuations
\be
\Delta t={\sqrt{ \langle \sigma_1^2 \rangle_R }\over r}\approx
\sqrt{ {32\pi G_{7}\over 12\pi^3L^3}}\,\sqrt{\ln(\gamma)}
=\sqrt{ {8G_4\over3\pi^2}}\sqrt{\ln(\gamma)}\approx {2\sqrt{2}t_{pl}\over \sqrt{3}\pi}\,
\sqrt{\ln(\gamma)}\,.
\label{eq:t3}
\nonumber\\
\ee


\paragraph{The case with n=4}


This is the 8 dimensional spacetime with 4 extra dimensions. 
The relevant two-point function is

\be
 G_{xxxx}(t,x,{\bf 0},\,t',x',0,0, m_1L,...,m_4L)\equiv A_4(\rho)+B_4(\rho)\,
\label{eq:G4}
\ee
with 
\ben
A_4={1\over 8\pi^4}&\Biggl[&
-{912\rho^6\over (\rho^2+\alpha_4^2)^6}+{902\alpha_4^2\rho^4\over (\rho^2+\alpha_4^2)^6}
+{201\alpha_4^4\rho^2\over 2(\rho^2+\alpha_4^2)^6}\nonumber\\
&&+{24\rho^8\over \alpha_4^2(\rho^2+\alpha_4^2)^6}
 -{12\alpha_4^6\over (\rho^2+\alpha_4^2)^6}
\Biggr]\,,
\een 
and
\ben
B_4&=&{1\over 8\pi^4} \,
\biggl[{240\rho^7\over (\rho^2+\alpha_4^2)^{13\over2}}
-{540\alpha_4^2\rho^5\over (\rho^2+\alpha_4^2)^{13\over2}}
+{90\alpha_4^4\rho^3\over (\rho^2+\alpha_4^2)^{13\over2}}
+{15\alpha_4^6\rho\over 4(\rho^2+\alpha_2^2)^{11\over2}}
\biggr]\times\,\nonumber\\
&&\quad \ln\left( \frac {\sqrt{\rho^{2} + \alpha_4^2} + \rho}{\sqrt{\rho^{2} + \alpha_4^{2}} - \rho}\right )^2\,.
\een
One finds after carrying out the integration
\ben
&&\int_{a}^{b} dx \int_{a}^{b} dx'\, G_{xxxx}^{R }(t,x,{\bf 0},\,t',x',{\bf 0})
\nonumber\\
&&\quad =
 {1\over4\pi^4L^4 }\prod_{i=1}^4{\sum_{m_i=1}^{\infty}}^{\prime}
\Bigg[
-{64\gamma^6\over 3(\sum_{i=1}^4\,m_i^2 )(\gamma^2+\sum_{i=1}^4\,m_i^2 )^4}\nonumber\\
&&\quad\quad
-{10\gamma^8\over 3(\sum_{i=1}^4\,m_i^2 )^2(\gamma^2+\sum_{i=1}^4\,m_i^2 )^4}
+\frac{32\gamma^4}{(\gamma^2+\sum_{i=1}^4\,m_i^2 )^4}
+\frac{15\gamma^2(\sum_{i=1}^4\,m_i^2 )}{(\gamma^2+\sum_{i=1}^4\,m_i^2 )^4}\nonumber\\
&&\quad
-\frac{48\gamma^5+
16( \sum_{i=1}^4\,m_i^2)\gamma^3
+3(\sum_{i=1}^4\,m_i^2 )^2\gamma}{2(\sum_{i=1}^4\,m_i^2 +\gamma^2)^{9/2}}
\ln\left({\sqrt{\sum_{i=1}^4\,m_i^2 +\gamma^2}+ \gamma}\over{\sqrt{\sum_{i=1}^4\,m_i^2 +\gamma^2}- \gamma} \right)\Bigg]\,.\nonumber\\
	\label{eq: series3}
\een
The summation is  seen to be dominated by the second term when $\gamma \gg 1$ 
and 
in that case the summation turns out to be approximated by an integral as
\be
-{5\over6\pi^4L^4 }\prod_{i=1}^4\int_{1/\gamma}^{\infty}\,dx_i\,
{1\over (\sum_{i=1}^5 x_i^2)^2(\sum_{i=1}^5 x_i^2+1)^4}\approx {5\over 6\pi^4L^4 }\ln(\gamma)\,.
\ee
Therefore, one obtains
\be
\langle \sigma_1^2 \rangle_R \approx {5r^2\over48\pi^4L^4 }\ln(\gamma)\,,
\ee
and
\be
\Delta t={\sqrt{ \langle \sigma_1^2 \rangle_R }\over r}
\approx {\sqrt{10}t_{pl}\over\sqrt{3}\pi^{3\over2}}\,
\sqrt{\ln(\gamma)}\,.
\label{eq:t4}
\ee

\paragraph{The case with n=5}


This is the 9 dimensional spacetime with 5 extra dimensions.  
One finds in this case that
\ben
 &&G_{xxxx}(t,x,{\bf 0},\,t',x',0,0, m_1L,...,m_2L)\nonumber\\
&&\quad ={1\over 16\pi^3}\biggl[
{24\rho^{10}\over\alpha_5^3(\rho^2+\alpha_5^2)^7}
{648\rho^8\over\alpha_5(\rho^2+\alpha_5^2)^7}
-{4536\alpha_5\rho^6\over (\rho^2+\alpha_5^2)^7}
+{2520\alpha_5^3\rho^4\over 2(\rho^2+\alpha_5^3)^7}
\biggr]\,,\nonumber\\
\label{eq:G5}
\een
and
\ben
&&\int_{a}^{b} dx \int_{a}^{b} dx'\, G_{xxxx}^{(1)R }(t,x,0,0,\,t',x',0,m_iL)
\nonumber\\
&&=-{1\over8\pi^3L^5 }\prod_{i=1}^5{\sum_{m_i=1}^{\infty}}^{\prime}\biggl(
{8\gamma^{10}\over (\sum_{i=1}^5m_i^2)^{5\over2}(\gamma^2+\sum_{i=1}^5m_i^2 )^5}
 +{48\gamma^8\over (\sum_{i=1}^5m_i^2)^{3\over2}(\gamma^2+\sum_{i=1}^5m_i^2 )^5}
\nonumber\\
&&\quad +{168\gamma^6\over (\sum_{i=1}^5m_i^2 )^{1\over2}(\gamma^2+\sum_{i=1}^5m_i^2 )^5}
\biggr)\,.
\een
 The  summation, to the leading order, can be approximated by integration
\be
-{1\over\pi^4L^5 }\prod_{i=1}^5\int_{1/\gamma}^{\infty}\,dx_i\,
{1\over (\sum_{i=1}^5 x_i^2)^{5\over2}(\sum_{i=1}^5 x_i^2+1)^5}\approx {1\over\pi^4L^5 }\ln(\gamma)\,, 
\ee
 as $\gamma\rightarrow \infty$.
Thus the mean deviation from the classical propagation time due to the lightcone fluctuations
\be
\Delta t={\sqrt{ \langle \sigma_1^2 \rangle_R }\over r}
\approx {2t_{pl}\over\pi^{3\over2}}\,
\sqrt{\ln(\gamma)}\,.
\label{eq:t5}
\ee


\paragraph{The case with n=6}


This is the 10 dimensional spacetime with 4 extra dimensions as motivated by superstring
theory. 
The relevant two-point function is given by

\be
 G_{xxxx}(t,x,{\bf 0},\,t',x',0,0,m_1L,...,m_6L)\equiv A_6(\rho)+B_6(\rho)\,,
\label{eq:G6}
\ee
where
\ben
A_6={1\over 16\pi^5}&\Biggl[&
-{15102\rho^6\over (\rho^2+\alpha_6^2)^7}
+{9102\alpha_6^2\rho^4\over (\rho^2+\alpha_6^2)^7}
+{4575\alpha_6^4\rho^2\over 2(\rho^2+\alpha_6^2)^7}\nonumber\\
&&+{816\rho^8\over \alpha_6^2(\rho^2+\alpha_6^2)^7}
 -{180\alpha_6^6\over (\rho^2+\alpha_6^2)^7}
+{48\rho^{10}\over \alpha_6^4(\rho^2+\alpha_6^2)^7}\Biggr]\,,
\een 
and
\ben
B_6&=&{1\over 16\pi^5} \,
\biggl[{4200\rho^7\over (\rho^2+\alpha_6^2)^{15\over2}}
-{6300\alpha_6^2\rho^5\over (\rho^2+\alpha_6^2)^{15\over2}}
+{1575\alpha_6^4\rho^3\over 2(\rho^2+\alpha_6^2)^{15\over2}}
+{105\alpha_6^6\rho\over 4(\rho^2+\alpha_6^2)^{11\over2}}
\biggr]\times\,\nonumber\\
&&\quad \ln\left( \frac {\sqrt{\rho^{2} + \alpha_6^2} + \rho}{\sqrt{\rho^{2} + \alpha_6^{2}} - \rho}\right )^2\,.
\een
One finds after carrying out the integration
\ben
&&\int_{a}^{b}dx \int_{a}^{b} dx'\, G_{xxxx}^{R }(t,x,0,0,\,t',x',0,m_iL) =\nonumber\\
&&\quad {1\over8\pi^5L^6 }\prod_{i=1}^6{\sum_{m_i=1}^{\infty}}^{\prime}
\Bigg[
-{424\gamma^6\over 3(\sum_{i=1}^6\,m_i^2 )(\gamma^2+\sum_{i=1}^6\,m_i^2 )^5}
-{68\gamma^8\over 3(\sum_{i=1}^6\,m_i^2 )^2(\gamma^2+\sum_{i=1}^6\,m_i^2 )^5}
\nonumber\\ &&\quad\quad
+\frac{390\gamma^4}{(\gamma^2+\sum_{i=1}^6\,m_i^2 )^5}
+\frac{195\gamma^2(\sum_{i=1}^6\,m_i^2 )}{(\gamma^2+\sum_{i=1}^6\,m_i^2 )^5}
-{4\gamma^{10}\over 3(\sum_{i=1}^6\,m_i^2 )^3(\gamma^2+\sum_{i=1}^6\,m_i^2 )^5}
\nonumber\\ &&\quad\quad
-\frac{400\gamma^5+
100( \sum_{i=1}^6\,m_i^2)\gamma^3
+15(\sum_{i=1}^4\,m_i^2 )^2\gamma}{2(\sum_{i=1}^4\,m_i^2 +\gamma^2)^{11/2}}
\ln\left({\sqrt{\sum_{i=1}^6\,m_i^2 +\gamma^2}+ \gamma}\over{\sqrt{\sum_{i=1}^6\,m_i^2 +\gamma^2}- \gamma} \right)\Bigg]\,.\nonumber\\
	\label{eq: series4}
\een
The summation is dominated by the second term when $\gamma \gg 1$
 and 
thus  is  approximated by an integral as
\be
-{1\over6\pi^5L^6 }\prod_{i=1}^6\int_{1/\gamma}^{\infty}\,dx_i\,
{1\over (\sum_{i=1}^6 x_i^2)^3(\sum_{i=1}^6 x_i^2+1)^5}\approx {1\over 6\pi^5L^6 }\ln(\gamma)\,.
\ee
Hence, we have 
\be
\langle \sigma_1^2 \rangle_R \approx {r^2\over48\pi^5L^6 }\ln(\gamma)\,,
\ee
and
\be
\Delta t={\sqrt{ \langle \sigma_1^2 \rangle_R }\over r}
\approx {\sqrt{2}t_{pl}\over\sqrt{3}\pi^2}\,
\sqrt{\ln(\gamma)}\,.
\label{eq:t6}
\ee


\paragraph{The case with n=7}


This is the 11 dimensional spacetime with 7 extra dimensions .  
The relevant two-point function can be shown to be
\ben
 &&G_{xxxx}(t,x,{\bf 0},\,t',x',0,0,m_1L,...,m_7L)\nonumber\\
&&\quad ={1\over 16\pi^5}\biggl[
{72\rho^{12}\over\alpha_7^5(\rho^2+\alpha_7^2)^8}
+{1056\rho^{10}\over\alpha_7^3(\rho^2+\alpha_7^2)^8}
{14256\rho^8\over\alpha_7(\rho^2+\alpha_7^2)^8}\nonumber\\
&&\quad\quad\quad\quad
-{66528\alpha_7\rho^6\over (\rho^2+\alpha_7^2)^8}
+{27720\alpha_7^3\rho^4\over 2(\rho^2+\alpha_7^3)^8}
\biggr]\,.\nonumber\\
\label{eq:G7}
\een
Following the same steps, one finds for $\gamma\gg 1$
\be
\langle \sigma_1^2\rangle_R\approx{5\over 16 \pi^5L^7}\prod_{i=1}^7\int_{1/\gamma}^{\infty}\,dx_i\,
{1\over (\sum_{i=1}^7 x_i^2)^{7\over2}(\sum_{i=1}^7 x_i^2+1)^6}\approx{5\over 16 \pi^5L^7}\ln(\gamma)\,,
\ee
and
\be
\Delta t={\sqrt{ \langle \sigma_1^2 \rangle_R }\over r}
\approx {\sqrt{10}t_{pl}\over\pi^4}\,
\sqrt{\ln(\gamma)}\,.
\label{eq:t7}
\ee

Thus we find that in all of these models in which there is more than one flat
compactified extra dimension, the lightcone fluctuation effect grows only
logarithmically with distance. Hence we are unable to derive any constraints
on these models. We have not been able to prove that this behavior holds for any
number of flat extra dimensions greater than one, but conjecture that this 
is the case.


\section{Parallel brane-worlds scenario}


In this section, we examine the lightcone fluctuations due to the presence of 
two 3+1 dimensional hyperplane boundaries, i.e., `` 3-branes'', living in extra dimensions 
separated from each other by some distance.  This framework is motivated by
the recent proposal to resolve the unnatural hierarchy between the weak and Planck scales
\cite{ADD}. 
In this scenario, four-dimensional particle theory, such as the Standard Model, 
 is confined to live in one of the branes, but gravity is free to propagate in 
the higher dimensional bulk. Therefore the bulk spacetime is dynamical, and the 3-branes can 
not be rigid, but must undergo quantum fluctuations in their positions, which we assume to be
order of $l_p$, the Planck length in higher dimensions. To be more specific,  we suppose one brane is located at the origin and the 
other at  $(0,0,0, z_1, z_2,..., z_n)$,  and we shall examine the effect of lightcone 
fluctuations due to gravitons in the bulk by looking at a light ray traveling parallel 
to one of the boundaries, in the $x-$axis, for example, but separated from it by a distance 
$z\sim l_p$. 
This feature is to reflect
 the quantum fluctuations in the position of the brane~\cite{MS99}. 
We consider both the Neumann  and Dirichlet boundary conditions
 for the graviton field at each brane hyperplane. 
Once we have the graviton two-point functions without any boundary, those  with two parallel
 plane boundaries can also be found by the method of image sum.
If gravitons satisfy the Dirichlet boundary condition, then the renormalized graviton
two-point function is given by the following multiple image sum
\ben
 &&G_{\mu\nu\rho\sigma}^{ R}(t, z_i,\,t',z_i')\nonumber\\
&&\quad
 =\prod_{i=1}^n{\sum_{m_i=-\infty}^{+\infty}}^{\prime}
G_{\mu\nu\rho\sigma}(t, z_i   , \,t',z_i' +2m_iL)
- \prod_{i=1}^n{\sum_{m_i=-\infty}^{+\infty}}\,
 G_{\mu\nu\rho\sigma}(t, z_i   , \,t',-z_i' +2m_iL)\nonumber\\
&& \quad
=\prod_{i=1}^n{\sum_{m_i=1}^{\infty}}\,\biggl[
 G_{\mu\nu\rho\sigma}(t, z_i   , \,t', 2m_iL+z_i')
-G_{\mu\nu\rho\sigma}(t, z_i   , \,t', 2m_iL-z_i')\nonumber\\
&&\quad\quad\quad\quad\quad
+ G_{\mu\nu\rho\sigma}(t, z_i   , \,t', -2m_iL+z_i')
-G_{\mu\nu\rho\sigma}(t, z_i   , \,t', -2m_iL-z_i')\biggr]\nonumber\\
&&\quad\quad\quad\quad\quad
-G_{\mu\nu\rho\sigma}(t, z_i   , \,t', -z_i')
\,\nonumber\\ 
\een
Again  the prime denotes omitting the $m_i=0$ term in the summation. 
For the  Neumann boundary, the renormalized graviton two-point function becomes
\ben
 &&G_{\mu\nu\rho\sigma}^{ R}(t, z_i,\,t',z_i')\nonumber\\
&&\quad
 =\prod_{i=1}^n{\sum_{m_i=-\infty}^{+\infty}}^{\prime}
G_{\mu\nu\rho\sigma}(t, z_i   , \,t',z_i' +2m_iL)
+\prod_{i=1}^n{\sum_{m_i=-\infty}^{+\infty}}\,
 G_{\mu\nu\rho\sigma}(t, z_i   , \,t',-z_i' +2m_iL)\nonumber\\
&&\quad
=\prod_{i=1}^n{\sum_{m_i=1}^{\infty}}\,\biggl[
 G_{\mu\nu\rho\sigma}(t, z_i   , \,t', 2m_iL+z_i')
+G_{\mu\nu\rho\sigma}(t, z_i   , \,t', 2m_iL-z_i')\nonumber\\
&&\quad\quad\quad\quad
+ G_{\mu\nu\rho\sigma}(t, z_i   , \,t', -2m_iL+z_i')
+G_{\mu\nu\rho\sigma}(t, z_i   , \,t', -2m_iL-z_i')\biggr]\nonumber\\
&&\quad\quad\quad\quad
+G_{\mu\nu\rho\sigma}(t, z_i   , \,t', -z_i')
\, .
\een

Let us examine a light ray propagating  along the
$x$-axis starting from $(a,0,0,z_1,...,z_n)$ to $(b,0,0,z_1,...,z_n)$. For simplicity, let
us assume that $z_1= \cdots =z_n=z\approx l_p/\sqrt{n}$. 
 Then the mean squared
 fluctuation in the geodesic interval function is
\be
\langle \sigma_1^2 \rangle = {1\over 8}(b-a)^2
\int_{a}^{b} dx \int_{a}^{b} dx'
\:\, G_{xxxx}^R(t,x,0,0,z_i,\,t',x',0,0, z_i). \,
\ee 


\subsection{ Five dimensional  theory}

Here we have one extra dimension, and the relevant two-point function is
\ben
G_{xxxx}(t,x,0,0,z_1, t',x',0,0,z_1')&=&
{1\over 4\pi^2}\Biggl[-{\rho^6\over(\rho^2+\Delta z_1 ^2)^3 |\Delta z_1|^3}
-{7\rho^4\over (\rho^2+\Delta z_1^2)^3 |\Delta z_1|}\nonumber\\
&&
+{9\rho^2|\Delta z_1|\over (\rho^2+\Delta z_1^2)^3}
- {|\Delta z_1|^3\over (\rho^2+\Delta z_1^2)^3}\Biggr]\,,
\een
where $\Delta z_1=z_1-z_1'$. We then have
\ben
&&\int_a^b\,dx\int_a^b\,dx'\,G_{xxxx}(t,x,0,0,z_1, t',x',0,0,z_1')\nonumber\\
&&\quad
={1\over \pi^2|\Delta z_1|}
\ln\bigl(1+{r^2\over \Delta z_1^2}\bigl)-{r^2\over 4\pi^2|\Delta z_1|^3}
-{r^2\over \pi^2(r^2+\Delta z_1^2) |\Delta z_1|}\nonumber\\
&&\quad\equiv f(\Delta z_1,r)\,.
\een
If gravitons satisfy the Dirichlet boundary conditions,  it   follows that
\ben
\langle \sigma_1^2 \rangle &=& {1\over 8}r^2
\int_{a}^{b} dx \int_{a}^{b} dx'
\:\, G_{xxxx}^R(t,x,0,0,z_i,\,t',x',0,z_i)\nonumber\\
&=&- {1\over 8}r^2f(2z,r)
+{1\over 8}r^2 \sum_{m=1}^{\infty}\,[2f(2mL,r)-f(2z-2mL,r)-f(2z+2mL,r)] \,.\nonumber\\
\een
Here we are interested in the case in which $r\gg z$ and $L\gg z$. 
One then finds that
\be
f(2z,r)\approx -{1\over L}{r^2\over 32\pi^2z^2}{L\over z}
=-{1\over L}{\gamma^2\over 32\pi^2}\biggl({L\over z}\biggr)^3\, ,
\ee
and
\ben
\sum_{m=1}^{\infty}\,&[&2f(2mL,r)-f(2z-2mL,r)-f(2z+2mL,r)]\nonumber\\
&&
 \approx {1\over 32\pi^2}
\sum_{m=1}^{\infty}\,\biggl[-{2r^2\over m^3L^3}+
{r^2\over (mL+z)^3}+{r^2\over (mL-z)^3}\biggr]
\nonumber\\
&&={1\over 32\pi^2L}
\sum_{m=1}^{\infty}\,\biggl[-{2\gamma^2\over m^3}+{\gamma^2\over (m-z/L)^3}
+{\gamma^2\over (m+z/L)^3}\biggr]
\nonumber\\
&&\approx {\gamma^2\over 32\pi^2L}
\sum_{m=1}^{\infty}\,\Biggl[{12\over m^5}\biggl({z\over L}\biggr)^2+{30\over m^7}
\biggl({z\over L} \biggr)^4
+O((z/L)^6)\Biggr]\nonumber\\
&&\approx {\gamma^2\over 32\pi^2L}
\sum_{m=1}^{\infty}\,{12\over m^5}\biggl({z\over L}\biggr)^2
={3\gamma^2\zeta(5)\over 8\pi^2L}\biggl({z\over L}\biggr)^2\,.
\een
Here $\zeta(5)$ is the Riemann-zeta function. Consequently, we find
\be
\langle \sigma_1^2 \rangle={\gamma^2r^2\over 8^2\pi^2L}\biggl[
{1\over 4}\biggl({L\over z}\biggl)^3+3\zeta(5)\biggl({z\over L}\biggl)^2\biggr]\,,
\ee
thus,
\be
\Delta t={\sqrt{ \langle \sigma_1^2 \rangle_R }\over r}
\approx \gamma\, t_{pl}\sqrt{{1\over 2\pi}\biggl[
{1\over 4}\biggl({L\over z}\biggl)^3+3\zeta(5)\biggl({z\over L}\biggl)^2\biggr]}
\approx {1\over2\sqrt{2\pi}}\biggl({L\over z}\biggl)^{3\over2}
\biggl({r\over L}\biggr)\,t_{pl}\,.
\label{eq:t8}
\ee

If instead of the Dirichlet boundary condition, gravitons are forced to satisfy the Neumann 
boundary condition, one finds  
\be
\langle \sigma_1^2 \rangle=-{\gamma^2r^2\over 8^2\pi^2L}\biggl[
{1\over 4}\biggl({L\over z}\biggl)^3+\zeta(3)+3\zeta(5)\biggl({z\over L}\biggl)^2\biggr]\,,
\ee
and
\be
\Delta t={\sqrt{ |\langle \sigma_1^2 \rangle_R| }\over r}
\approx \gamma\, t_{pl}\sqrt{{1\over 2\pi}\biggl[
{1\over 4}\biggl({L\over z}\biggl)^3+\zeta(3)+3\zeta(5)\biggl({z\over L}\biggl)^2\biggr]}
\approx {1\over2\sqrt{2\pi}}\biggl({L\over z}\biggl)^{3\over2}\biggl({r\over L}\biggr)\,t_{pl}\,.
\label{eq:t9}
\ee
Here we find that the choice of different  boundary conditions has very little effect on the
growth of the lightcone fluctuations as long as $L\gg z$, although the fluctuations are 
slightly larger in the case of the  Neumann boundary condition. Comparing this $\Delta t $ with 
that in the case of periodical compactification Eq.~(\ref{eq:tkk}), one can see that the effect of 
light 
cone fluctuation here is larger for a given size of extra dimensions, $L$.  Another very
distinctive feature is that here $\Delta t $ increases as $L$ increases provided that  $L$ and
 $z$ are 
independent of each other, which is in
sharp contrast with the case of periodical compactification. 
As before, the 
fluctuation in the flight time of pulses, $\Delta t$, can be applied to the successive 
wave crests of a plane wave. This leads to a broadening of spectral lines from luminous
sources. In case of periodical compactification, we used the data  
from gamma ray bursters to derive a very strong lower bound on $L$.  
In the present framework, however, 
we have an intrinsic lower  bound $L\geq z$.  Recall that $z$ is the Planck 
length in 
higher dimensions which may be much larger than that in 4 dimensions. 
In the picture of Ref.~\cite{ADD}, 
$z$  and $L$ are not independent, and 
may be expressed in terms of the Planck mass in $4+n$ dimensions, 
$M_{p}^{(4+n)}$, as
\be
l_{p}=2\times 10^{-17}\biggl( {1Tev\over M_{p}^{(4+n)}}\biggr)\,{\rm cm}\,,
\ee 
\be
L=10^{30/n-17}\,\biggl( {1Tev\over M_{p}^{(4+n)}}\biggr)^{1+2/n}\,{\rm cm}\,.
\ee
One has for the ratio $L/z$
\be
{L\over z}={L\over l_{p}}=0.5\times 10^{30/n}\biggl( {1Tev\over M_{p}^{(4+n)}}\biggr)^{2/n}\,.
\label{eq:ratio}
\ee
Express Eq.~(\ref{eq:t8}) or Eq.~(\ref{eq:t9})  as
\be
\Delta t \approx {1\over2\sqrt{2\pi}}\biggl({L\over z}\biggl)^{1\over2}
\biggl({r\over z}\biggr)\,t_{pl}\,,
\label{eq:t10}
\ee
or equivalently,  
\be
\Delta t \approx 10^{15}\biggl( {1Tev\over M_{p}^{(4+n)}}\biggr)^{1/n-1}
{r\over 10^{-17} cm}\,t_{pl}=10^{32}\,\biggl( {1Tev\over M_{p}^{(4+n)}}
\biggr)^{1/n-1}\biggl({ r\over 1\,cm}\biggr)\,t_{pl}\,.
\ee
Let $M_{p}^{(4+n)}$ be $\sim 1$Tev.  Then for an astronomical  source of 
cosmological distance, such as gamma-ray bursters with redshift $\sim 1$, 
or $r\sim 10^{28}$cm, we get the  following estimate 
for the $n=1$ model
\be
\Delta t\sim 10^{60}\times 10^{-44}s=10^{16}s.
\ee
That is far too large, so that the $n=1$ model can be ruled out completely.


\subsection{Higher dimensional theories}


We shall only study the 6 dimensional case in some detail, but  shall list  results 
for other cases.
 It is not difficult to see that the relevant two-point function  can be obtained from 
Eq.~(\ref{eq:G2}) 
by replacing $\alpha_2$ there with $\sum_{i=1}^2 \Delta z_i^2$. Therefore one easily 
finds from Eq.~(\ref{eq:series2}) 
\ben
&&\int_a^b\,dx\int_a^b\,dx'\,G_{xxxx}(t,x,0,0,z_1, z_2,\, t',x',0,0,z_1', z_2')\equiv f_2(\Delta z_1,\Delta z_2,r)\nonumber\\
&&\quad ={1\over4\pi^3}\Bigg[
\frac{r^2(\Delta z_1^2+\Delta z_2^2)}{(r^2+\Delta z_1^2+\Delta z_2^2)^3}
+\frac{10r^4}{3(r^2+\Delta z_1^2+\Delta z_2^2 )^3}\nonumber\\
&&\quad\quad
-{8r^6\over 3(\Delta z_1^2+\Delta z_2^2 )(r^2+\Delta z_1^2+\Delta z_2^2 )^3}
\nonumber\\
&&\quad \quad-\frac{8r^5+
4( \Delta z_1^2+\Delta z_2^2)r^3+(\Delta z_1^2+\Delta z_2^2 )^2r}
{2(\Delta z_1^2+\Delta z_2^2+r^2)^{7/2}}
\ln\left({\sqrt{\Delta z_1^2+\Delta z_2^2 +r^2}+ r}
\over{\sqrt{\Delta z_1^2+\Delta z_2^2 +r^2}- r} \right)\Bigg]\,. \nonumber\\
 \een
 Then it follows that
\ben
\langle \sigma_1^2 \rangle &=& {1\over 8}r^2
\int_{a}^{b} dx \int_{a}^{b} dx'
\:\, G_{xxxx}^R(t,x,0,0,z,z,\,t',x',0,z,z)\nonumber\\
&=&- {1\over 8}r^2f(2z,2z,r)
+{1\over 8}r^2 \sum_{m_1=1}^{\infty}\,\sum_{m_2=1}^{\infty}\, [2f(2m_1L,2m_2L,r)\nonumber\\
&&
-f(2z-2m_1L,2z-m_2L,r)-f(2z+2m_1L,2z+2m_2L,r)] \,.\nonumber\\
\label{eq:sigmaD}
\een
Here we are interested in the case when $r\gg z$ and $L\gg z$. One finds that
\be
f(2z,2z,r)\approx -{1\over 6\pi^3L^2}\biggl({L\over z}\biggr)^2
-{4\over \pi^3L^2}{\ln(r/z)\over\gamma^2}\,,
\ee
and
\ben
g(r,z,L)&\equiv& \sum_{m_1=1}^{\infty}\sum_{m_2=1}^{\infty} [\,2f(2m_1L,2m_2L,r)
-f(2z-2m_1L,2z-2m_2L,r)\nonumber\\
&&-f(2z+2m_1L,2m_2L,r)\,]\nonumber\\
&\approx& {1\over 4\pi^3L^2}
\sum_{m_1=1}^{\infty}\sum_{m_2=1}^{\infty}\,\Biggl(\,-{4\gamma^6\over
 3(m_1^2+m_2^2)
(\gamma+m_1^2+m_2^2)^3}\nonumber\\
&&
+{2\gamma^6\over 3[(m_1+z/L)^2+(m_2+z/L)^2][\gamma^2+(m_1+z/L)^2+
(m_2+z/L)^2 ]^3}
\nonumber\\&& 
+{2\gamma^6\over 3[(m_1-z/L)^2+(m_2-z/L)^2][\gamma^2+(m_1-z/L)^2+
(m_2-z/L)^2 ]^3}
\Biggr)\, . \nonumber\\
\een
If we assume the travel distance $r$ is much larger than the size of the extra 
dimensions $L$, i.e $\gamma \gg 1$, $g(r,z,L)$ can be approximated by integrals 
as follows
\ben
g(r,z,L)&\approx&
-{1\over6\pi^3L^2 }\int_{1/\gamma}^{\infty}\,dx_1\,\int_{1/\gamma}^{\infty}\,dx_2\,
{2\over (x_1^2+x_2^2)(x_1^2+x_2^2+1)^3}\nonumber\\
&&
+{1\over6\pi^3L^2 }\int_{{1\over\gamma}(1-z/L)}^{\infty}\,dx_1\,
\int_{{1\over\gamma}(1-z/L) }^{\infty}\,dx_2\,{1\over (x_1^2+x_2^2)(x_1^2+x_2^2+1)^3}
\nonumber\\
&&
+{1\over6\pi^3L^2 }\int_{{1\over\gamma}(1+z/L)}^{\infty}\,dx_1\,
\int_{{1\over\gamma}(1+z/L) }^{\infty}\,dx_2\,{1\over (x_1^2+x_2^2)(x_1^2+x_2^2+1)^3}
\nonumber\\ &\approx& {1\over6\pi^3L^2 }\Biggl[2\ln(\gamma)+\ln\bigg({1-z/L\over \gamma}\biggr)
+\ln\bigg({1+z/L\over \gamma}\biggr)\Biggr]\nonumber\\
&=&{1\over6\pi^3L^2 }\ln\biggl[1-(z/L)^2\biggr]\,.
\nonumber\\
\een
Therefore,
\ben
\langle \sigma_1^2 \rangle &=& {1\over 8}r^2
\int_{a}^{b} dx \int_{a}^{b} dx'
\:\, G_{xxxx}^R(t,x,0,0,z,z,\,t',x',0,z,z)\approx -{r^2\over 8}
f(2z,2z,r)\nonumber\\
&\approx& {r^2\over 48\pi^3L^2}\biggl({L\over z}\biggr)^2
+{2r^2\over 3\pi^3L^2}{\ln(r/z)\over\gamma^2}\,.
\een
Hence, 
\be
\Delta t={\sqrt{ \langle \sigma_1^2 \rangle_R }\over r}\approx
\sqrt{2\over3\pi^2}\biggl({L\over z}\biggr)\,.
\ee
Notice here that the leading term does not depend on the travel distance $r$ and moreover 
the next order corrections decrease as $r$ increases. Recall that
$\gamma=r/L$. Interestingly,  the lightcone
 fluctuation grows as the compactification scale increases.

However, if we change to the Neumann boundary condition, the behavior will be different as 
we can see from the following analysis.
Let us note that in this case
\ben
\langle \sigma_1^2 \rangle &=& {1\over 8}r^2
\int_{a}^{b} dx \int_{a}^{b} dx'
\:\, G_{xxxx}^R(t,x,0,0,z,z,\,t',x',0,z,z)\nonumber\\
&=& {1\over 8}r^2f(2z,2z,r)
+{1\over 8}r^2 \sum_{m_1=1}^{\infty}\,\sum_{m_2=1}^{\infty}\, [2f(2m_1L,2m_2L,r)\nonumber\\
&&
+f(2z-2m_1L,2z-m_2L,r)+f(2z+2m_1L,2z+2m_2L,r)] \,.\nonumber\\
\een
Notice the sign changes in the above expression as compared to 
Eq~(\ref{eq:sigmaD}). 
Then it follows that
\ben  
\langle \sigma_1^2 \rangle &\approx&  -{r^2\over 48\pi^3L^2}\biggl(
{L\over z}\biggr)^2
-{2r^2\over 3\pi^3L^2}{\ln(r/z)\over\gamma^2}+
{r^2\over48\pi^3L^2 }\Biggl[2\ln(\gamma)-
\ln\bigg({1-z^2/L^2\over \gamma^2}\biggr)
\Biggr]\nonumber\\
&\approx& {r^2\over 12\pi^3L^2}\ln(\gamma)-{r^2\over 48\pi^3L^2}\biggl(
{L\over z}\biggr)^2\,,
\een
and
\be
\Delta t={\sqrt{ \langle \sigma_1^2 \rangle_R }\over r}\approx
\sqrt{\biggl|{2\over3\pi^2}\biggl({L\over z}\biggr)^2-{8\over 3\pi^2}\ln(r/L)\biggr| }\,\,t_{pl}\,.
\ee
Here we have two different contributing terms to the mean deviation in time;  one  is 
independent of $r$ but increases as $L$ increases, the other grows logarithmically as $r$
increase or $L$ decreases. 

Similarly, we have calculated cases up to $n=7$ as motivated by string/M theory. 
The results are (all in units of $t_{pl}$, the Planck time in four dimensions):

 For n=3, 
\ben
  &&\Delta t\approx\sqrt{1\over6 \pi^3}\biggl({L\over z}\biggr)^{3\over2}\,,
 \quad \quad \quad \quad\quad \quad \quad \quad \quad \quad\ \,\,
{\rm  for \  Dirichlet \ boundary \ condition}\nonumber\\
&&
\Delta t\approx
\sqrt{\biggl|{1\over6\pi^3}\biggl({L\over z}\biggr)^3
-{2\over 3\pi^3}\ln(r/L)\biggr|}\,, 
\quad \quad \quad \quad {\rm for \ Neumann \ boundary \ condition.}\nonumber\\
\een

For n=4, 
\ben
  &&\Delta t\approx\sqrt{5\over12 \pi^4}\biggl({L\over z}\biggr)^{2}\,,
 \quad \quad \quad \quad\quad \quad \quad \quad \quad \quad\ \,\,
{\rm  for \  Dirichlet \ boundary \ condition}\nonumber\\
&&
\Delta t\approx
\sqrt{\biggl|{5\over12\pi^4}\biggl({L\over z}\biggr)^4-{5\over 3\pi^4}\ln(r/L)
\biggr| }\,, 
\quad \quad \quad \quad {\rm for \ Neumann \ boundary \ condition.}\nonumber\\
\een

For n=5, 
\ben
  &&\Delta t\approx{1\over4 \pi^2}\biggl({L\over z}\biggr)^{5\over2}\,,
 \quad \quad \quad \quad\quad \quad\quad \quad \quad \quad \quad \quad\,\,
{\rm  for \  Dirichlet \ boundary \ condition}\nonumber\\
&&
\Delta t\approx
\sqrt{\biggl|{1\over16\pi^4}\biggl({L\over z}\biggr)^5-{1\over 4\pi^4}\ln(r/L)
\biggr| }\,, 
\quad \quad \quad \quad {\rm for \ Neumann \ boundary \ condition.}\nonumber\\
\een

For n=6, 
\ben
  &&\Delta t\approx{1\over8\sqrt{3} \pi^{5\over2}}\biggl({L\over z}\biggr)^{3}\,,
 \quad \quad \quad \quad\quad \quad \quad \quad \quad \quad\ \,\,\,
{\rm  for \  Dirichlet \ boundary \ condition}\nonumber\\
&&
\Delta t\approx{1\over 8}
\sqrt{\biggl|{1\over3\pi^5}\biggl({L\over z}\biggr)^6-{4\over 3\pi^5}\ln(r/L)
\biggr| }\,, 
\quad \quad \quad \quad {\rm for \ Neumann \ boundary \ condition.}\nonumber\\
\label{eq:tn6}
\een

For n=7, 
\ben
  &&\Delta t\approx{1\over8}\sqrt{5\over2 \pi^5}\biggl({L\over z}\biggr)^{7\over2}\,,
 \quad \quad \quad \quad\quad \quad \quad \quad \quad \quad\ \,\,
{\rm  for \  Dirichlet \ boundary \ condition}\nonumber\\
&&
\Delta t\approx{1\over 8}
\sqrt{\biggl |{5\over2\pi^5}\biggl({L\over z}\biggr)^7-{10\over \pi^5}\ln(r/L)
\biggr | }\,, 
\quad \quad \quad \quad \, {\rm for \ Neumann \ boundary \ condition.}\nonumber\\
\een

 A few comments are now in order for all the cases we have examined: 
 First, to the 
leading order $\Delta t$ does not grow as $r$ increases for the Dirichlet boundary 
condition.  Second, 
$\Delta t $ behaves differently for different boundary conditions, but the 
logarithmic 
dependence associated with the Neumann boundary condition can be neglected as 
far as
observation is concerned because logarithmic growth is extremely small. Third, 
our results 
seem to suggest a leading order behavior of $(L/z)^{n/2}$ for any dimensions. 
Finally, our results
are quite different from those obtained by Ref.~\cite{CSEMN} where only the 
contribution of the $F$ term
 was considered and only one dominant term in the sums  computed. Our results
are much smaller than those obtained by these authors because of cancellations
among the various terms.

We wish to gain an understanding of how large the effect of lightcone 
fluctuations can be in these
higher dimensions and whether they might be observable. 

   From Eq.~(\ref{eq:ratio})  and the above results, one finds
\be
\Delta t\sim 10^{15}t_{pl}\biggl( {1Tev\over M_{p}^{(4+n)}}
                                                  \biggr)\,.
  \label{eq:del_t_6D}
\ee
This  result reveals that the smaller is $M_p^{(4+n)}$, the larger is the 
effect of 
lightcone fluctuations. According to Ref.~\cite{ADD}, $M_p^{(4+n)}$ may as low 
as the order of one Tev. This leads
to 
\be
\Delta t\sim 10^{-29}s\,.
\ee
This is a tiny effect by conventional standards. However if we note that the 
above mean time 
deviation is equivalent to an uncertainty in position 
\be
 \Delta x\sim 10^{-21}\,{\rm }m
\ee
 and the 
operation of gravity-wave interferometers is based upon the detection of minute 
changes in 
the positions of some test masses (relative to the position of a beam 
splitter),  we can see that this effect might be testable in the next 
LIGO/VIRGO generation of gravity-wave 
interferometers \cite{LIGO,VIRGO}. It constitutes an additional source of noise 
due to quantum gravity. 
Currently, the sensitivity of these gravity-wave interferometers has already 
reached the order of $10^{-19}$m \cite{AA}. Note that Amelino-Camelia \cite{GAC}
and Ng and van Dam \cite{NV} have also proposed rather different 
quantum gravity effects
which might also be detectable by laser interferometers.     

A more complete discussion of the observability of the $\Delta t$ given by
Eq.~(\ref{eq:del_t_6D}) should involve a calculation of the correlation time,
analogous to that performed in Sect.~\ref{sec:corr} for the five dimensional
Kaluza-Klein model. This calculation has not yet been performed. However,
it is reasonable to guess that the result will be of the order of or less
than Tev scale which characterizes this model. Recall that in the five
dimensional compactified model, the correlation time was found to be much
smaller than the compactification scale when the travel distance is large.
If this guess is correct, then the correlation time is much smaller than
$\Delta t$ itself.


\section{ Discussion and conclusions}


In this paper, we have examined the effects of compactified extra dimensions
upon the propagation of light in the uncompactified dimensions. There are 
nontrivial effects that arise from quantum fluctuations of the  
gravitational field, induced by the compactification. These effects take the 
form of lightcone fluctuations, variations in the flight times of pulses between
a source and a detector. The crucial quantity describing these fluctuations is
$\Delta t$, the expected variation in arrival times of two successive pulses
which are separated by more than a correlation time. In Sect.~\ref{sec:del_t},
we gave a derivation of the formula for $\Delta t$ using the geodesic deviation 
equation. This derivation allowed us to discuss issues of gauge and Lorentz 
invariance. In particular, it demonstrates that $\Delta t$ is gauge invariant.
All of the subsequent explicit calculations are performed in the 
transverse-tracefree gauge. 

As a prelude, we found the graviton two-point function in this gauge in 
Minkowski spacetime of arbitrary dimension. The  two-point function in
a higher dimensional flat compactified spacetime is given as an image sum. 
We then calculated $\Delta t$ in the five-dimensional Kaluza-Klein model,
five-dimensional flat spacetime with one periodic spatial dimension.
We found that $\Delta t$ grows linearly with increasing distance  between
the source and the detector, and is inversely proportional to the 
compactification length, $L$. This result differs from the square root
dependence on distance that was found in four-dimensional flat spacetime 
with one periodic spatial dimension \cite{YUF}. This demonstrates that
lightcone fluctuation effects are rather model-dependent. We also calculated 
the correlation time in the five-dimensional model and found that it
is typically small compared to the compactification scale $L$. This allows us
to place very tight constraints on the parameters of this model. 

We favor the viewpoint that the lightcone fluctuation effects should vanish
only in the limit that $L \rightarrow \infty$. If one adopts this view,
then the five-dimensional Kaluza-Klein model can be ruled out. Data from
gamma ray burst sources imply a lower bound on $L$ which is larger than
upper bounds obtained from other considerations. However, another logical
possibility is that lightcone fluctuation effects happen to vanish
($\langle \sigma_1^2\rangle =0$) at the present compactification scale.
Even if one adopts this viewpoint, one still obtains very strong constraints
on the fractional change in $L$ which can have occurred over a cosmological time
scale. This in turn tightly constrains any five-dimensional Kaluza-Klein
cosmology. 

We also examined an alternative  five-dimensional model (the brane worlds 
scenario) in which gravitons satisfy Dirichlet or Neumann boundary conditions
on a pair of parallel four-dimensional branes (one of which represents our 
world). In this model, we again find that the lightcone fluctuations are
so large that the model can be ruled out.

We next turned our attention to models with more than one extra dimension.
In the case of two or more flat, periodically compactified dimensions,
we found that $\Delta t$ grows only logarithmically with distance, and
that no constraints may be placed on these models. In the case of the
brane worlds scenario with more than one extra dimension, we found that
$\Delta t$ approaches a constant which can be of the order of $10^{-29}s$. 
This would produce a source of noise in laser interferometer detectors 
of gravity waves, which may eventually be within their range of sensitivity.

In summary, compactified extra dimensions have the possibility to produce 
observable effects by enhancing the quantum fluctuations of the gravitational
field. These effects might be used either to place constraints on theories
with extra dimensions, or else possibly eventually to provide positive 
evidence for the existence of  extra dimensions. Although our results for
more than one flat compactified extra dimension are too weak for either
of these purposes, the required calculations for models with curved extra 
dimensions have not yet been performed. Another model which has not yet
been examined in this context is the Randall-Sundrum model \cite{RS99}, with an
uncompactified fifth dimension. In this latter model, propagating graviton 
modes are effectively confined within a finite volume in the fifth dimension,
so one might expect nonzero lightcone fluctuations.

\begin{acknowledgments}
We would like to thank Ken Olum for helpful discussions. 
This work was supported in part by the National Science Foundation under
 Grant PHY-9800965.
\end{acknowledgments}

\section*{Appendix}
\setcounter{equation}{0}
\renewcommand{\theequation}{A\arabic{equation}}


\section{Graviton two-point functions in spacetime with arbitrary number of extra dimensions}


Here we evaluate the functions $D^n(x.x')$, $F^n_{ij}(x,x')$ and 
$H^n_{ijkl}(x,x')$ defined 
in Eqs.~(\ref{eq:Ffunc}), (\ref{eq:Dfunc}) and (\ref{eq:Hfunc}), respectively.  Once these functions are given, the graviton
two point functions are easy to obtain. 
Define
\be
R=|{\bf x}-{\bf x}'|, \quad \Delta t=t-t',\quad k=|{\bf k}|=\omega \,,
\ee
and assume $n$  extra dimensions, then
\ben
D^{n}(x,x')&=&
{Re\over{(2\pi)^{3+n}}}\int\, {d^{(3+n)}{\bf k}\over{2 \omega}}e^{i{\bf k} \cdot({\bf x}-{\bf x'})}e^{-i\omega(t-t')} \,\nonumber\\
&=&{Re\over{2(2\pi)^{3+n}}}\int_0^{\infty}\,k^{n+1}e^{-ik\Delta t}\,dk
\int_0^{\pi}\, d \theta_1\,\sin^{1+n}\theta_1 e^{i k R \cos\theta_1}
\nonumber\\&&\quad \times
\int_0^{\pi}\, d \theta_2\,\sin^{n}\theta_2...
\int_0^{\pi}\, d \theta_{n+1}\,\sin\theta_{n+1}
\int_0^{2\pi}\,d\theta_{n+2}\, \nonumber\\
&=&{a_n\,Re\over{2(2\pi)^{3+n}}}\,\int_0^{\infty}\,k^{n+1}e^{-ik\Delta t}\,dk
\int_{-1}^{1}\,e^{ikRx}(1-x^2)^{n/2}\,dx\,\nonumber\\
&=&{a_n\,Re\over{(2\pi)^{3+n}}}\,\int_0^{\infty}\,k^{n+1}e^{-ik\Delta t}\,dk
\int_{0}^{1}\,(1-x^2)^{n/2}\cos(kRx)\,dx\,\nonumber\\
&=&{a_n\sqrt{\pi}2^{{n+1\over 2}}\Gamma({n\over2}+1)\,Re\over{2(2\pi)^{3+n}}}\,
{1\over R^{n+1\over 2}}\int_0^{\infty}\,k^{n+1\over2}J_{n+1\over2}(kR)
e^{-ik\Delta t}\,dk\nonumber\\
&=&{a_n\sqrt{\pi}2^{{n+1\over 2}}\Gamma({n\over2}+1)\,Re\over{2(2\pi)^{3+n}}}\,
{1\over R^{n+1\over 2}}\lim_{\alpha\rightarrow 0^++i\Delta t}
\int_0^{\infty}\,k^{n+1\over2}J_{n+1\over2}(kR)e^{-\alpha k}\,dk\nonumber\\
&=&{a_n 2^n \Gamma({n\over2} +1)^2\over (2\pi)^{3+n}}
{1\over (R^2-\Delta t^2)^{n/2+1}}
={\Gamma({n\over2}+1)\over 4\pi^{ n+4\over2}}
{1\over (R^2-\Delta t^2)^{n/2+1}}\,.  \nonumber\\
\een
Here we have defined
\be
a_n=\int_0^{\pi}\, d \theta_2\,\sin^{n}\theta_2...
\int_0^{\pi}\, d \theta_{n+1}\,\sin\theta_{n+1}
\int_0^{2\pi}\,d\theta_{n+2}={2\pi^{ {n\over2}+1}\over\Gamma({n\over2}+1)}\,,
\ee 
and used 
\be
\int_{0}^{1}\,\cos(kRx)(1-x^2)^{n/2}\,dx=
{\sqrt{\pi}\over 2}\Gamma({n\over2}+1)\biggl({2\over kR}\biggr)^{n+1\over 2}J_{n+1\over2}(kR)\,,
\ee
and
\be
\int_0^{\infty}e^{-\alpha x}J_{\nu}(\beta x)x^{\nu}\,dx=
{(2\beta)^{\nu}\Gamma(\nu+1/2)
\over\sqrt{\pi}(\alpha^2+\beta^2)^{\nu+{1\over2}}}\,, \quad\quad Re \nu> -1/2\,.
\ee
When  $n$ is odd,  $D^n(x,x')$ should 
be taken to be zero when $R^2<\Delta t^2$.

Let us now turn our attention to the calculation of $F_{ij}$ and $H_{jikl}$. We find
\ben
F_{ij}^n(x,x')&=&{Re\over{(2\pi)^{3+n}}}\int\, d^{3+n}{\bf k}{k_ik_j
\over{2 \omega^3}}e^{i{\bf k} \cdot({\bf x}-{\bf x'})}e^{-i\omega(t-t')} 
\nonumber\\
&=&{Re\over{2(2\pi)^{3+n}}}\partial_i\partial_j'\int_0^{\infty}\,k^{n-1}
e^{-ik\Delta t}\,dk
\int_0^{\pi}\, d \theta_1\,\sin^{1+n}\theta_1 e^{i k R \cos\theta_1}
\nonumber\\&&\quad \times
\int_0^{\pi}\, d \theta_2\,\sin^{n}\theta_2...
\int_0^{\pi}\, d \theta_{n+1}\,\sin\theta_{n+1}
\int_0^{2\pi}\,d\theta_{n+2}\, \nonumber\\
&=&{a_n\,Re\over{(2\pi)^{3+n}}}\,\partial_i\partial_j'\int_0^{\infty}\,k^{n-1}
e^{-ik\Delta t}\,dk
\int_{0}^{1}\,(1-x^2)^{n/2}\cos(kRx)\,dx\,\nonumber\\
&=&{a_n\sqrt{\pi}2^{{n+1\over 2}}\Gamma({n\over2}+1)\,
Re\over{2(2\pi)^{3+n}}}\,\partial_i\partial_j'\left(
{1\over R^{n+1\over 2}}\int_0^{\infty}\,k^{n-3\over2}J_{n+1\over2}(kR)
e^{-ik\Delta t}\,dk\right)\nonumber\\
&=&{a_n\sqrt{\pi}2^{{n+1\over 2}}\Gamma({n\over2}+1)\over{2(2\pi)^{3+n}}}\,
\partial_i\partial_j'\left(
{Re\over R^{n+1\over 2}}\int_0^{\infty}\,k^{n-3\over2}J_{n+1\over2}(kR)
e^{-ik\Delta t}\,dk\right)\nonumber\\
&=&{Re\over 2(2\pi)^{3+n\over2}}\,\partial_i\partial_j'\left(
{1\over R^{n+1\over 2}}\int_0^{\infty}\,k^{n-3\over2}J_{n+1\over2}(kR)
e^{-ik\Delta t}\,dk\right)\nonumber\\
&=&{Re\over 2(2\pi)^{3+n\over2}}\,\partial_i\partial_j'\biggl(
{n-1\over R^2}{1\over R^{n-1\over 2}}\int_0^{\infty}\,k^{n-5\over2}
J_{n-1\over2}(kR)e^{-ik\Delta t}\,dk
\nonumber\\&&\quad-{1\over R^{n+1\over 2}}\int_0^{\infty}\,k^{n-3\over2}
J_{n-3\over2}(kR)e^{-ik\Delta t}\,dk\,\biggr)\,,
\label{eq:Fn}
\een
where we have utilized a recursive formula for Bessel functions
\be
zJ_{\nu-1}(z)+zJ_{\nu+1}(z)=2\nu J_{\nu}(z)\,.
\ee
Similarly, one finds that
\ben
H_{ijkl}^n(x,x')&&={Re\over{(2\pi)^{3+n}}}\int\, d^{3+n}{\bf k}{k_i k_j k_k k_l\over{2 \omega^5}}e^{i{\bf k} \cdot({\bf x}-{\bf x'})}e^{-i\omega(t-t')} \nonumber\\
&&={Re\over 2(2\pi)^{3+n\over2}}\,\partial_i\partial_j'\partial_k\partial_l\biggl(
{n-1\over R^2}{1\over R^{n-1\over 2}}\int_0^{\infty}\,k^{n-9\over2}J_{n-1\over2}(kR)e^{-ik\Delta t}\,dk
\nonumber\\&&\quad-{1\over R^2}{1\over R^{n+1\over 2}}\int_0^{\infty}\,k^{n-7\over2}J_{n-3\over2}(kR)
e^{-ik\Delta t}\,dk\,\biggr)\,.
\label{eq:Hn}
\een
To proceed further with the calculation, we need to deal with the cases when $n$ is odd
or even separately.
 

\subsection{The case of odd n}


 Assume $n=2m+1$ and define 
\be
S(m)={Re\over R^{m+1}}\,\int_0^{\infty}\,k^{m-1}J_{m+1}(kR)e^{-ik\Delta t}\,dk\,,\quad m\geq 0\,,
\ee
\ben
T(m-1)\,&=&{Re\over R^{m+1}}\,\int_0^{\infty}\,k^{m-1}J_{m-1}(kR)e^{-ik\Delta t}\,dk\nonumber\\
&=&{Re\over R^{m+1}}\,\lim_{\alpha\rightarrow 0^++i\Delta t}\int_0^{\infty}\,k^{m-1}J_{m-1}(kR)
e^{-alpha k}\,dk\nonumber\\
&=&{2^{m-1}\Gamma(m-1/2)\over\sqrt{\pi}}{\sqrt{R^2-\Delta t^2}\over R^2(R^2-\Delta t^2)^m}\,,\nonumber\\
&=&{(2m-1)!!\over (2m-1)} {\sqrt{R^2-\Delta t^2}\over R^2(R^2-\Delta t^2)^m}\,,\quad m\geq 1\,,
\een
where we have appealed to integral (6.623.1) in Ref.~\cite{GR1}.The above result holds for 
$R^2>\Delta t^2$, and $T(m-1)$ is zero when $R^2<\Delta t^2$.
Then it follows from Eq~(\ref{eq:Fn}) that 
\be
F^{2m+1}_{ij}={1\over 2(2\pi)^{m+2}}\,\partial_i\partial_j'\biggl(S(m)\biggr)\,,
                                               \label{eq:F}
\ee
and
\be
S(m)={2m\over R^2}S(m-1)-T(m-1)
\label{eq:Recursive1}
\ee

Using the recursive relation Eq~(\ref{eq:Recursive1}),
we can show that 
\ben
S(m)\,&=&{(2m)!!\over R^{2m}}S(0)-\sum_{k=1}^m{(2m)!!\over (2k)!!}{T(k-1)\over R^{2m-2k}}\nonumber\\
&=&{(2m)!!\over R^{2m}}S(0)\left[1-\sum_{k=1}^m{(2k-1)!!\over (2k)!!(2k-1)}
{R^{2k}\over (R^2-\Delta t^2)^k}\right]\nonumber\\
&=&\,-{(2m)!!\over R^{2m}}S(0)\,\sum_{k=0}^m{(2k+1)!!\over (2k)!!(2k+1)(2k-1)}
{R^{2k}\over (R^2-\Delta t^2)^k}\,.
\een
Here 
\be
S(0)=\left\{\begin{array}{ll} 
                 {\sqrt{R^2-\Delta t^2}\over R^2} &\mbox{ for}\, R^2>\Delta t^2\,,\\
                 0 &\mbox{ for}\, R^2<\Delta t^2\,,        
	    \end{array}\right.
\ee
as reported in Ref.~\cite{YUF2}.  For the sake of completeness, we give its 
derivation below
\begin{eqnarray}
S(0)&=& {Re\over R}\lim_{\alpha\rightarrow 0^++i\Delta t}\int_0^{\infty}\,{J_1(kR)\over k}
e^{-\alpha k}dk
    = {1\over R}\int_0^{\infty}\,{J_1(kR)\cos(k\Delta t)\over k}\,dk
\,\nonumber\\
&=& 
\left\{\begin{array}{ll}
             {1\over R}\cos\biggl(\arcsin(\Delta t/R)\biggr)={\sqrt{R^2-\Delta t^2}\over R^2}
                 &\mbox{ for}\, R^2>\Delta t^2\,,\\
             0 &\mbox{ for}\, R^2<\Delta t^2\,.        
	\end{array}\right.
\end{eqnarray}

If we define 
\be
Q(m)={Re\over R^{m+1}}\,\int_0^{\infty}\,k^{m-3}J_{m+1}(kR)
e^{-ik\Delta t}\,dk\,,\quad m\geq 0\,,
\ee
then it is easy to see that
\be
H_{ijkl}^{2m+1}={1\over 2(2\pi)^{m+2}}\,\partial_i\partial_j'
\partial_k\partial_l'\biggl(Q(m)\biggr)\,,
\ee
and 
\be
Q(m)={2m\over R^2}Q(m-1)-{1\over R^2}S(m-2)\,.
\label{eq:Recursive2}
\ee
The above equation applies for $m\geq 2$. To use it to get a general expression,
 we need $Q(0)$, which is given by 
\ben
Q(0)&=&{1\over R}\int_0^{\infty}\,
{1\over k^3}J_1(kR)\cos(k\Delta t)\,dk\,\nonumber\\
&=&
\lim_{\beta\rightarrow 0}\, {1\over R}\int_0^{\infty}\,
{k^{-1}\over (k^2+\beta^2)}J_1(kR)\cos(k\Delta t)\,dk\,\nonumber\\
&=&\lim_{\beta\rightarrow 0}\, {1\over R}{e^{-\beta\Delta t}I_1(\beta R)\over \beta^2}
={1\over 2\beta}-{1\over 2\Delta t}\,\,\nonumber\\
\een
This leads to a vanishing $H_{ijkl}$. 
We next  need  $Q(1)$, which can be calculated, using integral (6.693.5) in Ref.~\cite{GR1},
  as follows
\ben
Q(1)&=&{1\over R^2}\int_0^{\infty}\, {J_2(Rk)\cos(\Delta t k)\over k^2}\,dk\nonumber\\
&=&
  \left\{\begin{array}{ll}
       {1\over R^2}\left[ {R\over 4}\cos(\arcsin(\Delta t/R))+{R\over 12}\cos(3\arcsin(\Delta t/R))\right]
	  &\mbox{ for}\, R^2>\Delta t^2\,,\\
      0 &\mbox{ for}\, R^2<\Delta t^2\,.        
   \end{array}\right.
\nonumber\\
&=&
   \left\{\begin{array}{ll}
           \biggl({1\over3}-{\Delta t^2\over 3R^2}\biggr){\sqrt{R^2-\Delta t^2}\over R^2}
                =\biggl({1\over3}-{\Delta t^2\over 3R^2}\biggr)S(0) &\mbox{ for}\, R^2>\Delta t^2\,,\\
          0 &\mbox{ for}\, R^2<\Delta t^2\,.
    \end{array}\right. \nonumber\\
\een
In the above calculation, we have made use of the following trigonometric relations
\be
 \cos(3x)=4\cos^3(x)-3\cos(x), \quad\quad \cos(\arcsin x)=\sqrt{1-x^2}.\nonumber\\
\ee
Therefore one finds, using the recursive relation Eq~(\ref{eq:Recursive2}),
\ben
Q(m)&=&{(2m)!!\over 2R^{2m-2}}Q(1)-{1\over R^2}\sum_{k=2}^m{(2m)!!\over (2k)!!}{S(k-2)\over R^{2m-2k}}\nonumber\\
&=&{(2m)!!\over R^{2m-2}}\Biggl[{1\over2}Q(1)\nonumber\\
&&
+\,\sum_{k=2}^m\sum_{j=0}^{k-2}
{(2j+1)!!\over 2k(2k-2)(2j)!!(2j+1)(2j-1)}
{R^{2j}\over (R^2-\Delta t^2)^j}S(0)\,\Biggr]\,.
\nonumber\\
\een
This  expression can be simplified  if we note that 
\be
\sum_{k=j+2}^m\, {1\over k(k-1)}=
\sum_{k=2}^m\, {1\over k(k-1)}-\sum_{k=2}^{j+1}\, {1\over k(k-1)}
={{ m-j-1}\over m(j+1)}\,, 
\ee
and
\ben
 &&\sum_{k=2}^m\sum_{j=0}^{k-2}
{(2j+1)!!\over 2k(2k-2)(2j)!!(2j+1)(2j-1)}
{R^{2j}\over (R^2-\Delta t^2)^j}S(0)\,\Biggr]\nonumber\\
&&\quad=\sum_{j=0}^{m-2}\sum_{k=j+2}^m
{(2j+1)!!\over 2k(2k-2)(2j)!!(2j+1)(2j-1)}
{R^{2j}\over (R^2-\Delta t^2)^j}S(0)\,\Biggr]\nonumber\\
&&\quad=\sum_{j=0}^{m-2}
{(m-j-1)(2j+1)!!\over 4m(j+1)(2j)!!(2j+1)(2j-1)}
{R^{2j}\over (R^2-\Delta t^2)^j}S(0)\,\Biggr]\nonumber\\
\een
So, we have in this case 
\be
D^{2m+1}=
  \left\{\begin{array}{ll}
            {(2m+1)!!\over 2(2\pi)^{m+2}}{1\over (R^2-\Delta t^2)^{m+{3\over2}}}\,,
                      &\mbox{ for}\, R^2>\Delta t^2\,,\\
             0 &\mbox{ for}\, R^2<\Delta t^2\,,        
   \end{array}\right.
\ee
\be
F^{2m+1}_{ij}=-{1\over 2(2\pi)^{m+2}}\,\partial_i\partial_j'\left(
{(2m)!!\over R^{2m}}S(0)\,\sum_{k=0}^m{(2k+1)!!\over (2k)!!(2k+1)(2k-1)}
{R^{2k}\over (R^2-\Delta t^2)^k}\right)\,,
\ee
and
\ben
H_{ijkl}^{2m+1}&=&{1\over 2(2\pi)^{m+2}}\,\partial_i\partial_j'\partial_k\partial_l'\biggl\{
 {(2m)!!\over R^{2m-2}}\biggl[{1\over2}Q(1)\nonumber\\
&&+\,\sum_{j=0}^{m-2}
{(m-j-1)(2j+1)!!\over 4m(j+1)(2j)!!(2j+1)(2j-1)}
{R^{2j}\over (R^2-\Delta t^2)^j}S(0)\,\Biggr]\,.\nonumber\\
\een


\subsection{The case of even n}


Let $n=2m$ with $m=1,2,3...$. The graviton two-point functions for $m=0$ corresponding
to the usual 4 dimensional spacetime have been given previously \cite{YUF}.
The analog of Eq.~(\ref{eq:F}) for this case is
\be
F^{2m}_{ij}={1\over 2(2\pi)^{m+{3\over2}}}\,
\partial_i\partial_j'\biggl(S(m-{1\over2})\biggr) \,.
\ee
Here
\be
S(m-1/2)={2m-1\over R^2}S(m-{3\over2})-T(m-{3\over2})\,.
\label{eq:Recursive3}
\ee
Using this recursive relation, we can express $S(m-1/2)$ in terms of S(1/2) which is
 calculated, by employing
\be
J_{n+{1\over2}}(z)=(-1)^n\,z^{n+{1\over2}}\sqrt{{2\over\pi}}{d^n\over(zdz)^n}\biggl(
{\sin z\over z}\,\biggr)\,,
\ee
 to be
\ben
S(1/2)&&={1\over R^{3\over2}}\int_0^{\infty}\,k^{-1/2}J_{3\over2}(Rk)\cos(\Delta t k)\,dk
\nonumber\\
&&=-\sqrt{{2\over\pi}}{1\over R^3}\int_0^{\infty}\,{d\over dk}\biggl({\sin (Rk)\over k}
\biggr)\cos(\Delta t k)\,dk\nonumber\\
&&=-\sqrt{{2\over\pi}}\left({1\over R^3}{\sin (Rk)\cos(\Delta t k)\over k}
\Bigg |_0^{\infty}+
{\Delta t\over R^3}\int_0^{\infty}\,{\sin (Rk)\sin(\Delta t k)\over k}\,dk
\,\right)\nonumber\\
&&=\sqrt{{2\over\pi}}\left({1\over R^2}-{\Delta t\over4 R^3}\ln\biggl({R+\Delta t\over R-\Delta t}\biggr)^2\,\right)\,.
\een
It then follows that
\ben
S(m-1/2)\,&&={(2m-1)!!\over R^{2m-2}}S(1/2)-\sum_{k=2}^m{(2m-1)!!\over (2k-1)!!}
{T(k-3/2)\over R^{2m-2k}}\nonumber\\
&&={(2m-1)!!\over R^{2m}}\sqrt{{2\over\pi}}\left[1-{\Delta t\over4 R}\ln\biggl({R-\Delta t\over R-\Delta t}\biggr)^2\,
-{1\over R^2}\sum_{k=2}^m{2^{k-2}\Gamma(k-1)\over (2k-1)!!}
{R^{2k}\over (R^2-\Delta t^2)^{k-1}}\right]\,.
\nonumber\\
\een
Similarly, one has for $H_{ijkl}^{2m}$
\be
H^{2m}_{ijkl}={1\over 2(2\pi)^{m+{3\over2}}}\,\partial_i\partial_j'\partial_k\partial_l'
\biggl(Q(m-{1\over2})\biggr)\,,
\ee
and
\be
Q(m-1/2)={2m-1\over R^2}Q(m-{3\over2})-{1\over R^2}S(m-{5\over2})\,.
\label{eq:Recursive4}
\ee

Now the calculation becomes a little tricky. First, let us note that 
$H_{ijkl}^0$ has 
already been given \cite{YUF} and the recursive relation 
Eq~(\ref{eq:Recursive4}) can only be applied
when $m\geq 3$. So, we need both $H_{ijkl}^2$ and $H_{ijkl}^4$ or $Q(1/2)$ 
and $Q(3/2)$ 
as our basis to use the recursive relation for a general expression. 
 Because there is an infrared divergence in the $Q(1/2)$
 integral, so, as we did in the 4 dimensional case,
we will introduce a regulator $\beta$ in the 
denominator of the integrand and then let $\beta$ approach 0 after the integration is performed. 
Noting that
\be
J_{3\over2}(z)=\sqrt{2\over\pi z}\biggl({\sin z\over z}-\cos z\biggr)\,,
\ee
we obtain
\ben
Q(1/2)&=&{1\over R^{3\over2}}\,\int_0^{\infty}\,k^{-5\over2}
J_{3\over2}(kR)\cos(kt)\,dk\,
\nonumber\\
&&=\sqrt{2\over\pi }\biggl({1\over R^3} \int_0^{\infty}\,{dk\over k^4} \sin kR\, \cos k\Delta t\,-{1\over R^2} \int_0^{\infty}\,{dk\over k^3} \cos kR\, \cos k\Delta t\,\biggr)
\nonumber\\
&&=\sqrt{2\over\pi }\lim_{\beta\rightarrow 0}\biggl(-{1\over R^3}
{1\over 2\beta}{\partial\over\partial\beta}
 \int_0^{\infty}\,{\sin kR \cos k\Delta t\over k^2+\beta^2}\,dk 
\,\nonumber\\\quad\quad
&&\quad
+{1\over R^2}{1\over 2\beta}{\partial\over\partial\beta} \int_0^{\infty}\,
{k\cos kR\cos k\Delta t\over k^2+\beta^2} \,dk 
\,\biggr)\,.
\nonumber\\
\een
We next use
\ben
\int_0^{\infty}\,{\sin(ax)\cos(bx)\over \beta^2+x^2}\,dx=&&
{1\over4\beta}e^{-a\beta}
\{e^{b\beta}Ei[\beta(a-b)]+e^{-b\beta}Ei[\beta(a+b)]\}\nonumber\\
&&-{1\over4\beta}e^{a\beta}
\{e^{b\beta}Ei[-\beta(a+b)]+e^{-b\beta}Ei[-\beta(a-b)]\}\,,
\een
\ben
\int_0^{\infty}\,{x\cos(ax)\cos(bx)\over \beta^2+x^2}\,dx=&&
-{1\over4}e^{-a\beta}
\{e^{b\beta}Ei[\beta(a-b)]+e^{-b\beta}Ei[\beta(a+b)]\}\nonumber\\
&&-{1\over4}e^{a\beta}
\{e^{b\beta}Ei[-\beta(a+b)]+e^{-b\beta}Ei[-\beta(a-b)]\}\,,
\een
where
${\rm Ei(x)}$ is the exponential-integral function, and the fact that, when $x$ is small, 
\be
{\rm Ei(x)}\approx \gamma +\ln|x|+x+{1\over 4}x^2+{1\over 18}x^3+O(x^4)\,,
\ee
where $\gamma$ is the Euler constant. After expanding $Q(1/2)$ around 
$\beta=0$ to the order of $\beta^2$, one finds
\ben
Q(1/2)&=&\lim_{\beta\rightarrow0}\sqrt{2\over \pi}
\Biggl({5\over18}-{1\over3}\gamma-{1\over3}\ln(\beta)
-{1\over6}\ln(R^2-\Delta t^2)\nonumber\\
&&
-{\Delta t^2\over6R^2}+{\Delta t\over8R}\biggl({\Delta t^2\over3 R^2}-1\biggl)
\ln\biggl({R+\Delta t\over R-\Delta t}\biggr)^2
\Biggr)\,.
\een
Note, however, that what we need is $H_{ijkl}$ which involves differentiation 
of $Q(1/2)$, 
therefore we can discard the constant and divergent terms in $Q(1/2)$ as far 
as $H_{ijkl}$
is concerned.
To calculate $Q(3/2)$, let us recall that
\be
Q(3/2)={3\over R^2}Q(1/2)-{1\over R^2}S(-1/2)
\ee
 and note that $S(-1/2)$ is given by $\sqrt{2/\pi}$ times Eq~(A19) in Ref.~\cite{YUF}
Thus, we have
\be
Q(3/2)=\sqrt{2\over \pi}\left(-{1\over6R^2}
-{\Delta t^2\over2R^4}+{\Delta t\over8R^3}\biggl({\Delta t^2\over R^2}-1\biggl)
\ln\biggl({R+\Delta t\over R-\Delta t}\biggr)^2\right)\,.
\ee
With $Q(3/2)$ at hand, it is easy to show that for an arbitrary $m\geq 3$
\ben
Q(m-1/2)&=&{(2m-1)!!\over 3R^{2m-4}}Q(3/2)-
{1\over R^2}\sum_{k=3}^m{(2m-1)!!\over (2k-1)!!}{S(k-5/2)\over R^{2m-2k}}\nonumber\\
&=&{(2m-1)!!\over R^{2m-4}}\biggl(\,{1\over 3}Q(3/2)-\sum_{k=3}^m\,
{1\over(2k-1)(2k-3)}S(1/2)
\nonumber\\
&& +{1\over R^4}\sqrt{2\over\pi}\sum_{k=3}^m\sum_{j=2}^{k-2}\,
{2^{j-2}\Gamma(j-1)\over (2j-1)!!(2k-1)(2k-3)}{R^{2j}\over (R^2-\Delta t^2)^{j-1}}\,
\biggr)\nonumber\\
&=&{(2m-1)!!\over R^{2m-4}}\biggl(\,{1\over 3}Q(3/2)-\sum_{k=3}^m\,
{1\over(2k-1)(2k-3)}S(1/2)
\nonumber\\
&& +{1\over R^4}\sqrt{2\over\pi}\sum_{j=2}^{m-2}\,
{(m-j-1)2^{j-2}\Gamma(j-1)\over (2m-1)(2j+1)!!}{R^{2j}\over (R^2-\Delta t^2)^{j-1}}\,
\biggr)\,.\nonumber\\
\een
Here in the last step, we have made use of the following results
\ben
\sum_{k=j+2}^m\, {1\over(2k-1)(2k-3)}&=&
\sum_{k=2}^m\, {1\over (2k-1)(2k-3)}-\sum_{k=2}^{j+1}\, {1\over (2k-1)(2k-3)}\nonumber\\
&=&{{ m-j-1}\over (2m-1)(2j+1)}\,,
\een
and 
\be
\sum_{k=3}^m\sum_{j=2}^{k-2}\,f(j)g(k)=\sum_{k=4}^m\sum_{j=2}^{k-2}\,f(j)g(k)
=\sum_{j=2}^{m-2}f(j)\sum_{k=2+j}^{m}\,g(k)\,.
\ee
Consequently, we obtain
\be
D^{2m}={2^m m!\over (2\pi)^{m+2}}{1\over (R^2-\Delta t^2)^{m+1}}\,,
\ee
\ben
F^{2m}_{ij}&=&{1\over (2\pi)^{m+2}}\,\partial_i\partial_j'\biggl\{
{(2m-1)!!\over R^{2m}}\biggl[1-{\Delta t\over4 R}\ln\biggl({R-\Delta t\over R-\Delta t}\biggr)^2\,\nonumber\\
&&
-{1\over R^2}\sum_{k=2}^m{2^{k-2}\Gamma(k-1)\over (2k-1)!!}
{R^{2k}\over (R^2-\Delta t^2)^{k-1}}\biggr]\biggr\}\,,
\een
and
\ben
H_{ijkl}^{2m}&=&{1\over (2\pi)^{m+2}}\,\partial_i\partial_j'\partial_k\partial_l'\biggl\{{(2m-1)!!\over R^{2m-4}}\biggl[\,
{\Delta t\over24R^3}\biggl({\Delta t^2\over R^2}-1\biggl)
\ln\biggl({R+\Delta t\over R-\Delta t}\biggr)^2
-{1\over18R^2}\nonumber\\
&&\quad\quad
-{\Delta t^2\over6R^4}
-\sum_{k=3}^m\,{1\over (2k-1)(2k-3)}\left({1\over R^2}-{\Delta t\over4 R^3}\ln\biggl({R+\Delta t\over R-\Delta t}\biggr)^2\,\right)
\nonumber\\
&&\quad \quad+{1\over R^4}\sum_{j=2}^{m-2}\,
{(m-j-1)2^{j-2}\Gamma(j-1)\over (2m-1)(2j+1)!!}{R^{2j}\over (R^2-\Delta t^2)^{j-1}}\,
\biggr)\,.\nonumber\\
\een


\begin{references}

\bibitem{ACF} Th. Kaluza, Sitzungsber. Preuss. Akad. Wiss., Phys. Math. Kl.,
996 (1921); O. Klein, Z. Phys. {\bf 37}, 895 (1926).   


\bibitem{ANT90} I. Antoniadis,  Phys. Lett. {\bf B246}, 377 (1994).

\bibitem{KS91} V.A. Kostelecky and S. Samuel, Phys. Lett. {\bf B270}, 21 (1991).

\bibitem{AB94} I. Antoniadis and K. Benakli,  Phys. Lett. {\bf B326}, 69 (1994).

\bibitem {ADD} N. Arkani-Hamed, S. Dimopoulos and G. Dvali,  Phys. Lett. {\bf B429}, 263 (1998);
 I. Antoniadis, N. Arkani-Hamed, S. Dimopoulos and G. Dvali,  Phys. Lett.  {\bf B436}, 247 (1998);  N. Arkani-Hamed, S. Dimopoulos and G. Dvali, Phys. Rev.  {\bf D59}, 086004  (1999). 

\bibitem{WLST} E. Witten, Nucl. Phys. {\bf B471}, 135 (1996);  J, Lykken, Phys. 
Rev.  {\bf D 54}, 3693 (1996);  G. Shiu and S.-H.H. Tye, 
 Phys. Rev. {\bf D 58}, 106007 (1998).

\bibitem{Colliders} J.L. Hewett, Phys. Rev. Lett. {\bf 82}, 4765 (1999);\\
  G.F. Giudice, R. Rattazzi and  
J.D. Wells,  Nucl. Phys. {\bf B544}, 3 (1999); \\
T.G. Rizzo,  Phys. Rev. {\bf D59},  115010 (1999); \\
 I. Antoniadis, K. Benakli, and  M. Quiros, {\it  Direct collider signatures of large extra dimensions},  hep-ph/9905311;\\ 
 T. G. Rizzo and  J. D. Wells, Phys. Rev. {\bf D61},  016007 (2000) . 

\bibitem{DN} A. Kehagias and  K. Sfetsos, {\it Deviations from the $1/r^2$ Newton law due to extra dimensions},  hep-ph/9905417; \\ 
E.G. Floratos and  G.K. Leontaris, Phys. Lett. {\bf B465}, 95 (1999),  hep-ph/9906238. 

\bibitem{GU} K.R. Dienes, E.Dudas, and T. Ghergheta, Phys. Lett. {\bf 436}, 55 (1998), 
hep-ph/9803466;\\
 Nucl. Phys. {\bf 537}, 47(1999), hep-ph/9806292; \\
 D. Chilencea and G.G. Ross, Phys. Lett. {\bf B442}, 165(1998),
hep-ph/9809217;\\
  S.A. Abel and S.F. King, Phys. Rev. {\bf D59}, 095010 (1999), hep-ph/9809467. \\
C.D. Carrone, Phys. Lett. {\bf B454}, 70 (1999), hep-ph/9902407.\\
 A. Delgado and Quiros, Nucl. Phys. {\bf B559}, 235 (1999), hep-ph/9903400. \\
 A.  Perez-Lorenzana and  R. N. Mohapatra, Nucl. Phys. {\bf B559}, 255 (1999),  hep-ph/9904504.\\
  H-C. Cheng, B. A. Dobrescu and C. T. Hill, {\it Gauge Coupling Unification with Extra Dimensions and Gravitational Scale Effects},   hep-ph/9906327.


 \bibitem{Cosmo} N. Kaloper and A. Linde, Phys. Rev. D{\bf 59}, 101303 (1999);\\ 
 D. Lyth, Phys. Lett., B{\bf 448}, 191(1999);\\
 G. Dvali, and S.-H. Tye, Phys. Lett., B{\bf 450}, 72(1999);\\
 C. Csaki, M. Graesser and J. Terning, Phys. Lett., B{\bf 456}, 16(1999);\\
 G. Dvali,  Phys. Lett., B{\bf 459}, 489(1999);\\
 N. Arkani-Hamed, S. Dimopoulos, N. Kaloper and John March-Russell,
 {\it Rapid Asymmetric Inflation and Early Cosmology in Theories with Sub-Millimeter
Dimensions},  hep-ph/9903224;\\
 A. Mazumdar, Phys. Lett. {\bf B469}, 55 (1999),  hep-ph/9902381;\\
 L. J. Hall and D. Smith, Phys. Rev. {\bf D60}, 085008 (1999),  hep-ph/9904267; \\
 A. Riotto, {\it D-branes, String Cosmology and Large Extra Dimensions},   hep-ph/9904485;  \\
 J. M. Cline, Phys. Rev.  {\bf D61}, 023513 (2000),  hep-ph/9904495;\\
 A. Melchiorri, F. Vernizzi, R. Durrer, and G. Veneziano, 
Phys. Rev. Lett. {\bf 83}, 4464 (1999), astro-ph/9905327.

\bibitem{YUF}H. Yu and L.H. Ford, Phys. Rev. D{\bf 60}, 084023 (1999), gr-qc/9904082.

\bibitem{YUF2} H. Yu and L.H. Ford, {\it Lightcone fluctuations in quantum 
gravity and extra dimensions},  gr-qc/9907037.

\bibitem{APCH} T. Appelquist and A. Chodos, Phys. Rev. Lett. {\bf 50}, 141 (1983); Phys. Rev. 
{\bf D28}, 772 (1983).

\bibitem{RRT} M.A. Rubin and B. Roth, Phys. Lett. {\bf 127B}, 55 (1983); K. Tsokos, Phys. Lett. 
{\bf 126B}, 451 (1983). 

\bibitem{Ford95} L.H. Ford, Phys. Rev.  {\bf D51}, 1692
               (1995), gr-qc/9410043. 

\bibitem{WF99} C.-H. Wu and L.H. Ford, Phys. Rev. D {\bf 60}, 104013 
(1999), gr-qc/9905012.

\bibitem{Ford96} L.H. Ford and N. F. Svaiter, Phys. Rev.  {\bf D54}, 2640
               (1996), gr-qc/9604052
\bibitem{AEMN} G. Amelino-Camelia, J. Ellis,  N.E. Mavromatos and D.V. Nanopoulos, Int. J. Mod. 
Phys.  {\bf A12},  607 (1997). 


\bibitem{EMN} J. Ellis, N.E. Mavromatos and D.V. Nanopoulos, 
{\it Quantum-Gravitational Diffusion and Stochastic Fluctuations in the Velocity of Light}, 
  gr-qc/9904068; gr-qc/9905048. 

\bibitem{KPW} E.W. Kolb, M.J. Perry, and T.P. Walker, Phys. Rev.  {\bf D33},
 869 (1986). 

\bibitem{JDB} J.D. Barrow, Phys. Rev.  {\bf D35}, 1805 (1987). 

\bibitem{MM} M. Maurette, Ann. Rev. Nucl. Sci. {\bf 26}, 319 (1976).

\bibitem{AIS} A.I. Shylakhter, Nature (London), {\bf 264}, 340 (1976).

Phys.  {\bf A12},  607 (1997). 


\bibitem{SANDERS} D.B. Sanders et al, Astr. Ap., {\bf 215}, L5 (1989).

\bibitem{LCP} J.C. Long, H.W. Chan, and J.C. Price, Nucl. Phys. {\bf B539}, 23 
  (1999), hep-ph/9805217. 

\bibitem{TP99} T. Piran. Phys. Rep. {\bf 314}, 575 (1999).

\bibitem{MR97} M. Rees , astro-ph/9701162.

\bibitem{ACEMNS}  G. Amelino-Camelia, J. Ellis,  N.E. Mavromatos, 
D.V. Nanopoulos and S. Sarkar, Nature {\bf 393}, 763 (1998).

\bibitem{BES} B.E. Schaefer, Phys. Rev. Lett., {\bf 82}, 4964 (1999).

\bibitem{SDB} S.D. Biller et al,  Phys. Rev. Lett., {\bf 83}, 2108 (1999).

\bibitem{PK} P. Kaaret, Astronomy and Astrophysics, {\bf 345}, L3 (1999).

\bibitem{EFMMN} J. Ellis et al, {\it Astrophysical probes of the constancy 
of the velocity of light }, astro-ph/9907340.

\bibitem{PNB} P.N. Bhat {\it et al.,} Nature(London) {\bf 359}, 217 (1992).

\bibitem{KH} K. Hurley {\it et al.,} Nature(London) {\bf 372}, 652 (1994).

\bibitem{MS} M. Sommer {\it et al.,} Astrophys. J. {\bf 422}, L63 (1994).

\bibitem{GRB980425} S.R. Kulkarni, {\it et al}. Nature {\bf 395} 663 (1998).

\bibitem{GRB990123} S.R. Kulkarni, {\it et al}. Nature {\bf 398} 389 (1999);
T.J. Galama, {\it et al}. Nature {\bf 398} 394 (1999); C. Akerhof, {\it et al}.
Nature {\bf 398} 400 (1999).


\bibitem{DeWitt63} B.S. DeWitt,  
in {\it Relativity, Groups and Topology}, eds. B.S. and C.M. de Witt (Gordon and Breach, 
New York, 1963).

\bibitem{MS99}  N.E. Mavromatos  and R. Szabo,  Phys. Rev.  {\bf D59}, 104018
               (1999). 

\bibitem{CSEMN}A. Campbell-Smith,  J. Ellis, N.E. Mavromatos and 
D.V. Nanopoulos,  Phys. Lett. {\bf B466}, 11 (1999),  hep-th/9907141.

\bibitem{LIGO} A. Abramovici et al.,  Science 256, 325(1992).
\bibitem{VIRGO} C. Bradaschia et al., Nucl. Instrum. Meth. {\bf 289A}, 518 (1990).
\bibitem{AA} A. Abramovici et al., Phys. Lett., {\bf 218A}, 157 (1996).


\bibitem{GAC} G. Amelino-Camelia, Nature {\bf 398}, 216 (1999); 
{\it On the Salecker-Wigner limit and the use of interferometers in space-time-foam studies},  gr-qc/9910023;
{ \it Are we at the dawn of quantum-gravity phenomenology?}, gr-qc/9910089.

\bibitem{NV} Y.J. Ng and H. van Dam,  Phys.Lett. {\bf B477}, 429 (2000);
gr-qc/9906003.

\bibitem{RS99} L. Randall and R. Sundrum, Phys. Rev. Lett. {\bf 83}, 4690
(1999).
 
\bibitem{GR1} I.S. Gradshteyn and I.M. Ryzhik, {\it Table of Integrals,
Series, and Products}, (Academic Press, New York, 1965).


\end{references}
\end{document}